\documentclass[prc,eqsecnum,amsmath,amsfonts,nofootinbib]{revtex4}
\usepackage{graphicx}
\usepackage{bm}
\usepackage{pstricks}        
\usepackage{mathtools}
\usepackage{subcaption}
\usepackage{siunitx}
\usepackage{times}
\usepackage{euler}
\usepackage{calc}
\usepackage{amsthm}
\usepackage{amsmath}
\usepackage{amssymb}
\usepackage{float}
\usepackage{physics}
\usepackage{tikz}
\usepackage{mathdots}
\usepackage{yhmath}
\usepackage{array}
\usepackage{multirow}
\usepackage{braket}
\usepackage{dcolumn}
\newcolumntype{d}[1]{D{.}{\cdot}{#1}}
\newcolumntype{.}[1]{D{.}{.}{3.#1}}
\usepackage{enumitem}
\usepackage[super]{nth}
\usepackage{framed}
\makeatletter
\newcommand*{\length}[1]{%
    \@tempcnta\z@%
    \@for\@tempa:=#1\do{\expandafter\ifx\@tempa\empty\empty\else\advance\@tempcnta\@ne\fi}%
    \the\@tempcnta%
}
\makeatother
\newcommand{\MeijerGIndex}[5][]{
  {}^{#1\length{#4}\length{#2}}_{#1\length{#2,#3}\length{#4,#5}}%
  }
\newcommand{\MeijerGPars}[5][]{%
  \left|\begin{array}{ll}#1#2;&#3\\#1#4;&#5\end{array}\right.%
  }
\newcommand{\MeijerG}[5]{\ensuremath{%
G\MeijerGIndex{#2}{#3}{#4}{#5}\left(#1\MeijerGPars{#2}{#3}{#4}{#5}\right)%
}}

\renewcommand{\ldots}{\ensuremath\langle\mathbf{L}\cdot\mathbf{S}\rangle}
\allowdisplaybreaks[1]

\newcommand{\etal}{\textit{et al.}}
\newtheorem{theorem}{Theorem}
\newtheorem{lemma}[theorem]{Lemma}
\newtheorem{corollary}[theorem]{Corollary}

\newtheorem{definition}{Definition}

\newcommand{\thmref}[1]{Theorem \ref{#1}}
\newcommand{\lemref}[1]{Lemma \ref{#1}}
\DeclareMathOperator*{\K}{K}

\newcommand{\ibar}{\ensuremath{\overline{I}}}
\newcommand{\sbar}{\ensuremath{\overline{S}}}
\newcommand{\shat}{\ensuremath{\hat{S}}}
\begin{document}
\hyphenation{Rijken}
\title{Novel method for evaluating the eigenvalues of the Heun differential equation with an application to the Breit equation}
%
\author{P.J. Rijken}
\affiliation{Independent Researcher, Houten \\
    The Netherlands}
\email{pieter.rijken@gmail.com}
\author{Th.A. Rijken}
\affiliation{Radboud University Nijmegen \\
	The Netherlands}
\begin{abstract}
\begin{description}
\item[Background] Eigenvalues of the Breit equation
\[
\left[\left(\bm{\alpha}_1\cdot{\bf p} +
\beta_1 m\right)_{\alpha \alpha'}\delta_{\beta\beta'} + 
\delta_{\alpha \alpha'}\left(-\bm{\alpha}_2\cdot{\bf p} +
\beta_2 M\right)_{\beta\beta'} -\frac{e^2}{r}\delta_{\alpha \alpha'}
\delta_{\beta\beta'}\right] \Psi_{\alpha'\beta'} = E\ \Psi_{\alpha \beta},
\]
in which only the static Coulomb potential is considered, have been found. Over the past
decades several authors have analyzed the Breit equation to obtain numerically or by
approximation an estimation of the energy levels. Various approaches have been used
and no determination of the energy levels currently exists that is directly based
on the second order Heun differential equation derived from the equation above.
In addition, currently no explicit expression exists for the expansion in $\alpha^{2}$
of the energy levels beyond the first order.

\item[Purpose] The aim of this work is to develop the theoretical framework to obtain such
an expansion for the eigenvalues valid to $\alpha^4$ and beyond in a future work. Besides
the calculation of the energy levels, the aim is to provide a method of calculation that
can be used to numerically calculate the energy levels for various spin states to high
accuracy. Here, the detailed discussion on the simple case, $^1S_{0}$, $m=M$, is used to
illustrate our approach.

\item[Method] From the Breit equation, we derive the corresponding second-order Heun
differential equation and continued fraction from which the eigenvalues can be determined
very accurately. We show that Miller’s backward recurrence algorithm is a convenient
tool to determine the normalized radial wave functions.\\
Next, we present a novel method based on the Green function method, which leads to a
semi-infinite determinant from which we are able to obtain the numerical values of the
eigenvalues by direct calculation. This provides us with two distinct methods for
obtaining the eigenvalues. The methods that are based on either the continued fraction or on
Green functions utilize different mathematical tools and serve as an important consistency
check for their validity. To validate our results, we compare them with known results in the
literature.

\item[Results] Using suitable numerical methods for the direct calculation of the continued
fraction and the semi-infinite determinant, we show that both methods are consistent within
25 digits of accuracy. We describe how the radial wave functions are efficiently calculated.
Moreover, we show that the correct energy levels for the Dirac equation follow from our
results by a suitable mapping of the variables.

\item[Conclusions] The results are in total agreement with earlier calculations found in the
literature and extend this by several digits of additional accuracy. The condition on
the determinant giving the energy levels provides a rich structure that is promising in
extending the results of this work.
\end{description}
\end{abstract}
\maketitle
\date{version of: \today}
\tableofcontents
\section{\label{sec:intro}Introduction and the Breit equation}                               
The Breit-equation \cite{Breit32} is since long known, see \cite{BS57},
to lead to a more singular second-order differential equation
than the in mathematical physics usual appearing (confluent)
hypergeometrical type, see \cite{AS70}. 
The latter has basically three regular singular 
points, but the Breit equations lead to a second-order differential
equation with basically four singular points. Such an equation is 
known in the literature as the Heun-equation \cite{Heun89,Ronv95}.

Recently, the Breit equation has been studied by Yamaguchi \cite{Yam12} and
even more recent by Tursunov \etal \cite{Tur24}.
Here, the standard method for solving the Dirac equation for the
Hydrogen atom is used. In these notes, the methods particular to the 
Heun equation are used. Central is the derivation of a three-term 
recurrence relation, which contains the so-called "access" parameter.
These recurrences are solved using the continued fractions (CF) construction.
We compare our results with those of existing literature \cite{Sco92,Yam01,Tur24}.
In a paper by Yamaguchi \cite{Yam12}, numerical values for the eigenvalues
are determined by matching series solutions at $\rho=0$ and at $\rho=\infty$.
Tursunov \etal \cite{Tur24} have studied the energy levels based on
a variational method.
This work contributes to this result by offering an alternative way
to calculate approximations for the eigenvalues by solving the CF.

We extend this further by rewriting the Heun differential equation
as an integral equation. This allows us to set up a semi-infinite
determinant from which the eigenvalues can be extracted. This analysis
also shows that the series expansion of the eigenvalues has non-analytical
behavior in the electro-magnetic fine-structure constant $\alpha^2$. This is
a new result and in contrast with existing literature \cite{Ish51a,Ish51b}.
The main contributions of this paper are to give an alternative expression
from which the eigenvalues can be determined, thereby building upon and
extending the results of \cite{Yam01,Yam12}. This provides a method
to numerically calculate the eigenvalues to a high level of accuracy.

This paper is organized as follows.
In section \ref{sec:breit} we start with a discussion of the Breit equation
and its radial equation. We present the equations for both the spin-singlet
and spin-triplet cases, and introduce a parametrization that captures
the essential characteristics.
Next, in section \ref{sec:expansion} we solve the radial equation by expanding
in associated Laguerre polynomials and establish a three-term recurrent
relation for the coefficients in the expansion. This leads to a CF equation
for the eigenvalues. We prove the convergence of the series expansion for the
radial eigenfunctions and present graphs for $rF(r)$ and $rK(r)$.
As a follow-up, section \ref{sec:breitsinglet} focuses on the spin-singlet
case and provides a numerical solution to the eigenvalues based on the
continued fraction. In addition, we determine values for the energies for
the first five levels and angular momenta. This is compared to values
already established in the literature \cite{Sco92,Yam01}.
The spin-triplet case is addressed in section \ref{sec:spintriplet}. Here,
numerical results have been obtained for all cases except for the spin-triplet
states with $J\neq L$. A model to summarize the main properties of the
radial equations, is provided and uses three variables $\tau$, $\lambda$,
and $\kappa$.
Section \ref{sec:green} is devoted to rewriting the Breit differential
equation as a homogeneous Fredholm integral equation of the second kind
using Green's function method.
Using this result, we show in section \ref{sec:determinant} how the integral
equation enables us to solve for the eigenvalues using a condition on a
semi-infinite dimensional determinant.
As an intermezzo, section \ref{sec:dirac} discusses the resemblance
between the Breit and Dirac equations and that the solution to the
latter can be recovered from the determinant. Comparing the two cases
and how the eigenvalues of the Dirac equation are contained in the
condition on the determinant provides a profound understanding of
important properties of the determinant that are useful in the next
sections.
A thorough analysis of the convergence of the determinant is given
in section \ref{sec:detconv}. A so-called converging factor is introduced,
which fixes the determinant and is fundamental in obtaining numerical
results.
Finally, in section \ref{sec:smally} we provide a numerical verification
of the determinant using different methods. In addition to the CF,
we use a direct method and derive a forward recursion relation.
For various values of $y$ and sizes of the determinant up to $1024^2$ we
calculate the eigenvalues and verify that the results are consistent.
In the appendices \ref{sec:imnproperties}-\ref{sec:shatproperties} we
have gathered useful definitions, properties, summations, expansions, and
derivations of intermediate results that we believe are convenient for
anyone building upon this work.

\section{\label{sec:breit}The spin-singlet state: Breit radial differential equation}                                     
Here, we repeat the formulation of the radial differential equation
as given in Ref.~\cite{Yam12}.
%
\subsection{Basic equation}                                     
\label{sec:2a}
The CM Breit equation for the electron and proton, mass $m$ and $M$ respectively,
interacting through the (static) Coulomb potential is given by
\begin{equation}
\left[(\bm{\alpha}_1\cdot{\bf p}_{op}+\beta_1 m)_{\mu\nu}\delta_{\rho\sigma}
 +\delta_{\mu\nu}(-\bm{\alpha}_2\cdot{\bf p}_{op}+\beta_2 M)_{\rho\sigma}
 -\frac{e^2}{r}\delta_{\mu\nu}\delta_{\rho\sigma}\right] \Psi_{\nu\sigma} =
 E \Psi_{\mu\rho},
\label{eq:1.1}
\end{equation}
where ${\bf p}_{op} = -i\bm{\nabla}$. 
Concentrating on the bound-state region, henceforth the CM momentum
 is imaginary and is denoted by $p_{CM}=iq$. The total energy $E$ is given by
\begin{equation}
 E = \sqrt{m^2-q^2}+\sqrt{M^2-q^2}.
\label{eq:1.2}\end{equation}
%
\subsection{\label{sec:2b}The spin-singlet case}                                     
Following \cite{Yam12} the radial component functions $F(r)$ and $K(r)$ are defined by
\begin{equation}
\begin{tabular}{cc|c}
$\alpha$ & $\beta$ & $\Psi_{\alpha\beta}$ \\ \hline
1 & 1 & \multirow{2}*{$F(r)\ket{{}^1(\ell)_\ell}$} \\
2 & 2 & \\ \hline
1 & 3 & \multirow{2}*{$i \left\{ G(r)\ket{{}^3(\ell+1)_\ell} + \tilde{G}(r)\ket{{}^3(\ell-1)_\ell} \right\}$}\\
2 & 4 & \\ \hline
3 & 1 & \multirow{2}*{$i \left\{ H(r)\ket{{}^3(\ell+1)_\ell} + \tilde{H}(r)\ket{{}^3(\ell-1)_\ell} \right\}$} \\
4 & 2 & \\ \hline
3 & 3 & \multirow{2}*{$K(r)\ket{{}^1(\ell)_\ell}$} \\
4 & 4 & 
\end{tabular}
\end{equation}
From the Breit equation~\eqref{eq:1.1} the following set of 
differential equations is derived for the radial wave functions
$F(r)$, $K(r)$, $G(r)$, $\widetilde{G}(r)$, $H(r)$
and $\widetilde{H}(r)$\footnote{This set is consistent with equations (2.10) in \cite{Sco92}
and corrects the minus sign in the first equation of (6) in \cite{Yam12}.}:
\begin{subequations}
\begin{eqnarray}
&& \left(E+\frac{\alpha}{r}-m-M\right) F = \sqrt{\frac{j+1}{2j+1}}
 \left(\frac{d}{dr}+\frac{j+2}{r}\right)(G+H) -\sqrt{\frac{j}{2j+1}}
 \left(\frac{d}{dr}-\frac{j-1}{r}\right)(\widetilde{G}+\widetilde{H}), \\
&& \sqrt{j}
 \left(\frac{d}{dr}+\frac{j+2}{r}\right)(G-H) +\sqrt{j+1}
 \left(\frac{d}{dr}-\frac{j-1}{r}\right) (\widetilde{G}-\widetilde{H}) = 0, \\
&& \left[E+\frac{\alpha}{r}-(m+M)\right] F = 
\left[E+\frac{\alpha}{r}+(m+M)\right] K, \\
&& \left[E+\frac{\alpha}{r}+(m-M)\right] H = 
 \sqrt{\frac{j+1}{2j+1}}\left(\frac{d}{dr}-\frac{j}{r}\right)(F+K)
 = \left[E+\frac{\alpha}{r}-(m-M)\right] G, \\
&& \left[E+\frac{\alpha}{r}+(m-M)\right] \widetilde{H} = 
 -\sqrt{\frac{j}{2j+1}}\left(\frac{d}{dr}+\frac{j+1}{r}\right)(F+K)
 = \left[E+\frac{\alpha}{r}-(m-M)\right] \widetilde{G}. 
\label{eq:1.3}\end{eqnarray}
\end{subequations}
Here $j=\ell$, and in anticipation of the cases where $j\neq\ell$ in the equations above
we have preferred $j$ over $\ell$. In \cite{Yam12} the following dimensionless parameters
are introduced
\begin{subequations}
\begin{eqnarray}
 \rho &=& 2 qr, \label{eq:1.4a}\\
 y &=& \frac{E}{2\alpha q} = \frac{\sqrt{M^2-q^2}+\sqrt{m^2-q^2}}{2\alpha q} 
 >0, \\
 \lambda &=& \frac{m+M+E}{2\alpha q} > 0, \\
 \nu &=& \frac{m+M-E}{2\alpha q} > 0.     
\label{eq:1.4}\end{eqnarray}
\end{subequations}
In \cite{Yam12} it is noted that if the eigenvalue $E$ is a proper 
relativistic generalization of the Schr\"{o}dinger case, $y$ and
$\lambda$ are large and of the order $1/\alpha^2$, whereas $\nu$ is of
the order unity.\\
For $\rho \rightarrow \infty$ ($r \rightarrow \infty$), all radial wave
functions should behave as
\begin{equation}
 F(\rho) \rightarrow \rho^n\left[1+{\cal O}(1/\rho)\right]\ e^{-\rho/2}.
\label{eq:1.5}\end{equation}
%
\subsection{The case of equal masses $\bm{ m=M}$}
For equal masses, like in the case of the positron, $M=m$, in the singlet case,
the equations simplify considerably. First
\begin{equation}
  H=G,\ \widetilde{H}=\widetilde{G},\ K=\frac{1-\nu\rho}{1+\lambda\rho} F.
\label{eq:1.6}\end{equation}
Secondly, the combination 
\begin{equation}
 F+K \equiv \widetilde{h}(\rho) \exp\left( -\frac{1}{2}\rho\right)
\label{eq:1.7}\end{equation}
satisfies the differential equation
\begin{equation}
 \widetilde{h}'' + \widetilde{h}' \left\{-1+\frac{2}{\rho} + \frac{1}{\rho (1+y\rho)}\right\} + \widetilde{h}\left\{
 \left(\frac{1}{2}\alpha^2 y - 1\right)\frac{1}{\rho} +
 \frac{1}{4} \frac{\alpha^2}{\rho^2}-\frac{j(j+1)}{\rho^2}
 + \frac{-\frac{1}{2}}{\rho (1+y\rho)}\right\} = 0.
\label{eq:1.8}\end{equation}
Next, we introduce $k(\rho)$ by 
\begin{equation}
 \widetilde{h} \equiv \rho^{\tau-1-s}\ k(\rho), \,\, \tau = \sqrt{1+j(j+1)}, \,\, s = \tau - \sqrt{\tau^2 - \frac{1}{4}\alpha^2}, \,\, \text{or}\,\, s(2\tau-s) = \frac{1}{4}\alpha^2
\label{eq:staudefinition}
\end{equation}
so that $s$ is of the order $\alpha^2$, and $k(\rho)$ satisfies the differential equation
\begin{equation}
 k'' + k' \left\{-1 + \frac{1 + 2\tau - 2s}{\rho} - \frac{y}{1+y\rho}\right\} + k\left\{
 \left(\frac{1}{2}\alpha^2 y - \tau + s \right)\frac{1}{\rho}
 + \frac{-y (\tau-1-s) - \frac{1}{2}}{\rho (1+y\rho)} \right\} = 0.
\label{eq:basicform}
\end{equation}
Changing to the independent variable $\rho \equiv -\rho'/y$ we obtain finally the standard form:
\begin{equation}
 k'' + k' \left\{\frac{1}{y} - \frac{1}{\rho'-1} + \frac{1 + 2\tau - 2s}{\rho'} \right\} + k\left\{
 \frac{\left( -\frac{1}{2}\alpha^2 y + \tau - s \right) \rho'
 + \frac{1}{2}\alpha^2 y - (1+y) (\tau-1-s) - \frac{3}{2} }{y \rho'(\rho'-1)}
 \right\} = 0.
\end{equation}
See \cite{Yam12}, equation (2.7). Rewriting \eqref{eq:1.8} to the standard form
of the confluent Heun equation (eq. 1-2.27, part B, \cite{Ronv95}):
Eq. 1-2.27 in \cite{Ronv95} Part B reads:
\begin{eqnarray}
&& k^{\prime\prime}(\rho') +
 k'(\rho')\left(4p +\frac{\delta}{\rho'-1}+\frac{\gamma}{\rho'}\right)
 +k(\rho')\ \frac{4p \beta \rho'-\sigma}{\rho'(\rho'-1)} = 0.                  
\label{eq:2.7}\end{eqnarray}
Matching \eqref{eq:2.7} gives
\begin{eqnarray}
 4p &\rightarrow& 1/y, \nonumber\\[0.5em]
 \gamma &\rightarrow& 1 + 2\tau - 2s, \nonumber\\[0.5em]
 \delta & \rightarrow& -1, \nonumber\\ [0.5em]
 4p\beta & \rightarrow& \left( -\frac{1}{2}\alpha^2 y + \tau - s \right)/y,
\nonumber\\[0.5em]
 \sigma &\rightarrow& \tau-1-s + \left(-\frac{1}{2}\alpha^2 y + \frac{1}{2} + \tau - s\right)/y.
 \label{eq:sigmasinglet}
\end{eqnarray}
With these definitions, we can make the identification:
\begin{equation}
z = \frac{1}{2}\alpha^2 y
\label{eq:zdefinition}
\end{equation}
For specific values of '$y$', solutions exist that are bounded at
$\rho'=0$, i.e. $k(\rho') < \infty$ when $\rho' \rightarrow 0$, and
that approach zero as $\rho' \rightarrow \infty$, i.e. $k(\rho') \rightarrow 0$
as $\rho' \rightarrow \infty$. These solutions are called {\bf Heun functions}
and are valid throughout the entire $\rho$-plane. The specific values
$y=y_n, n=0,1,2,...$ are the eigenvalues of the equation. For convenience we
define $z_n = \frac{1}{2} \alpha^2  y_n$.\\

The differential equation satisfied by $F+K$ reads
\begin{equation}
(F+K)'' + \left(\frac{1+2\tau}{\rho} - \frac{y}{1+y\rho}\right) (F+K)' + \left(\frac{(1+\tau-s)(1-\tau+s)}{\rho^2} + \frac{z_N}{\rho} - \frac{1}{4}\right) (F+K) = 0
\end{equation}
Near $\rho=0$ it possesses a regular solution with exponent $\tau-1-s$, i.e. $F+K \sim \rho^{\tau-1-s}$.

In the vicinity of $\rho=\infty$ the equation has no regular solution. Following
Ince \cite{Ince56} section 17.5, a normal solution can be constructed with behavior
\begin{equation}
F+K \sim \rho^{z_N-1}\,e^{-\frac{1}{2}\rho}
\end{equation}
This leads to a condition on the behavior of $F+K$ for $\rho\rightarrow\infty$.
Boundary conditions on the radial wave function $F+K$ are:
\begin{subequations}
\begin{eqnarray}
\lim_{\rho\rightarrow 0} \rho (F+K) &=& 0, \label{eq:boundarycond1}\\
\int_0^\infty d\rho \rho^2\left|F+K\right|^2 &=& \left(\frac{2}{z}\, \frac{E}{2m}\, \frac{1}{2a_0}\right)^3, \label{eq:boundarycond2}\\
\lim_{\rho\rightarrow\infty} \left|F+K\right| &\sim& \rho^{z_N-1}\,e^{-\frac{1}{2}\rho}\left( a_0 + \frac{a_1}{\rho} + \frac{a_2}{\rho^2} + ... \right).
\label{eq:boundarycond3}
\end{eqnarray}
\end{subequations}
In deriving \eqref{eq:boundarycond2} in terms of the modified radial variable
$\rho$ we have used \eqref{eq:1.4} and rewritten $m\ \alpha=1/a_0$ where $a_0$
is the Bohr radius in units $\hbar=c=1$. This matches the numerical analysis in
\cite{Sco92} where radial components are expressed in units of $2a_0$.\\

These conditions stem from the requirements: (a) existence of the energy operator,
(b) total probability is one, and (c) existence of a bound state and existence of a
normal solution. Because of the additional (regular) singularity at $\rho=-1/y$, in
general, the Frobenius solution at $\rho=0$ has a radius of convergence of $1/y$. When
each of the two solutions at $\rho=0$ is analytically continued beyond this point, it
can be expressed as a linear combination of the two normal solutions at $\rho=\infty$.
The consequence is that, in general, it will be impossible to satisfy both the boundary
conditions (a) and (c) at the same time.\\

Fortunately, for specific values of $y$ (the eigenvalues) a solution can be constructed
that is valid in the entire $\rho$-plane except for the singularities at $0$, $\-1/y$,
and $\infty$. These solutions are known as Heun functions.

\section{\label{sec:expansion}Expansion in Associated Laguerre Polynomials}
To construct the solutions, we expand them in terms of the solutions 
of the {\bf confluent hypergeometric equation}, corresponding to the 
singularities at 0 and $\infty$. To this end, we follow Ref.~\cite{Ronv95}
part B equation 2.3.35. Rewriting equation \eqref{eq:2.7} we obtain
\begin{eqnarray}
&& \left[ (\rho'-1)\left\{\rho'\frac{d^2}{d\rho^{\prime 2}} + 
 4p\ \rho'\frac{d}{d\rho'} +\gamma \frac{d}{d\rho'}\right\}
 + \delta \rho'\frac{d}{d\rho'} + 4p\beta \rho'\right]\ k(\rho') =
 \sigma\ k(\rho'). 
\label{eq:2.9}\end{eqnarray}
Here, the operators occur
\begin{eqnarray}
 M_0 &\equiv& \rho'\frac{d^2}{d\rho^{\prime 2}} + 
 4p\ \rho'\frac{d}{d\rho'} +\gamma \frac{d}{d\rho'}, \nonumber\\  
 M_1 & \equiv& \delta \rho'\frac{d}{d\rho'} + 4p\beta \rho'.
\label{eq:2.10}\end{eqnarray}
We note that $M_0$ corresponds to the Kummer equation with singularities at 
0 and $\infty$. The eigenfunctions $v(\rho')$ are the {\bf confluent 
hypergeometric functions} $_1\!F_1$:
\begin{eqnarray}
 M_0\ v(\rho') &=& 4p \kappa v(\rho') \Rightarrow v(\rho') =\  
 \frac{1}{\Gamma(\gamma)} {}_1\!F_1(-\kappa; \gamma; -4p \rho'),
\label{eq:2.11}\end{eqnarray}
where $\kappa$ is the parametrization of the expansion. We choose:
\begin{eqnarray}
 \kappa &\equiv& \kappa_n = n + \nu,\ \ n=0,1,.... \nonumber\\
 v(\rho') &\equiv& v_n(\rho') =\ {}_1\!F_1(-n-\nu;\gamma; -4p \rho').
\label{eq:2.12}\end{eqnarray}
Utilizing well-known relations and identities \cite{AS70} for $_1\!F_1(...)$
we obtain the results:
\begin{eqnarray}
 M_1 v_n(\rho') &=& (\beta-\delta)(n+\nu)\ v_{n-1}(\rho') 
 +\left[(\delta-2\beta)(n+\nu)-\beta\gamma\right]\ v_n(\rho')
 +\beta (\gamma+n+\nu)\ v_{n+1}(\rho'),\ \ {\rm and} \nonumber\\[3mm]
 (\rho'-1) M_0\ v_n(\rho') &=& 4p (n+\nu)(\rho'-1)\ v_n(\rho')
 \nonumber\\ &=&
 (n+\nu)^2\ v_{n-1}(\rho')-(n+\nu)(4p+2n+2\nu+\gamma)\ v_n(\rho')
 +(n+\nu)(\gamma+n+\nu)\ v_{n+1}(\rho').
\label{eq:2.13}\end{eqnarray}
Expanding $k(\rho')$ in $v_n(\rho')$ as 
$k(\rho') = \sum_{n=-\infty}^\infty g_n\ v_n(\rho')$ leads to:
\begin{eqnarray}
&& \sum_{n=-\infty}^\infty \left[(n+1+\nu)(\beta-\delta+n+1+\nu)\ g_{n+1}
 +\left(\vphantom{\frac{A}{A}} (n+\nu)(-4p-\gamma+\delta-2\beta-2n-2\nu)
 -\beta \gamma-\sigma\right)\ g_n  
 \right.\nonumber\\ && \left. \vphantom{\frac{A}{A}}
 +(\gamma+n-1+\nu)(\beta+n-1+\nu)\ g_{n-1}
 \right]\ v_n(\rho') =0.
\label{eq:2.14}\end{eqnarray}
We arrive at the three-term recurrence relation:
\begin{equation}
 C_n\ g_{n+1} + B_n\ g_n + A_n\ g_{n-1} = 0, 
\label{eq:heunttrr}
\end{equation}
where
\begin{subequations}
\label{eq:heunabcgen}
\begin{eqnarray}
 A_n &=& (\gamma+n-1+\nu)(\beta+n-1+\nu), \\
 B_n &=& (n+\nu)(-4p-\gamma+\delta-2\beta-2n-2\nu)-\beta\gamma-\sigma, \\ 
 C_n &=& (n+1+\nu)(\beta-\delta+n+1+\nu).
\end{eqnarray}
\end{subequations}
Expressing equations \eqref{eq:heunabcgen} in terms of the variables $z$ and $s$ we get:
\begin{subequations}
\label{eq:heunabc}
\begin{eqnarray}
A_n &=& (n + 2\tau - 2s) (n - 1 - z + \tau - s) \\
B_n &=& -n \left( 2n + \frac{1}{y} + 2(1 + 2\tau - 2s - z) \right) - (1 + 2\tau - 2s) \left(\frac{1}{2y} - z + \tau - s\right) + \frac{z}{y} - (\tau - 1 - s + \lambda) \\
C_n &=& (n + 1) (n + 2 - z + \tau - s )
\end{eqnarray}
\end{subequations}
Converting the above recurrence relation to a continued fraction (CF) gives:
\begin{eqnarray}
 \frac{g_n}{g_{n-1}} &=& \frac{-A_n}{B_n+C_n\ \frac{g_{n+1}}{g_n} }
 = \frac{-A_n/C_n}{B_n/C_n + \frac{g_{n+1}}{g_n} }.
\label{eq:ctdfraction}
\end{eqnarray}
The energy $E$ is given by
\begin{equation}
\frac{E}{2m} = \frac{1}{\sqrt{1 + \frac{\alpha^2}{4 z^2}}}
\label{eq:energylevels}
\end{equation}
%
\subsection{Series expansion of the eigenfunctions}
In the parametrization of $\kappa_n$, the value of $\nu$ is undetermined. A convenient
choice is to take $\nu=0$. With this choice the functions $v_n(\rho')$ become (Laguerre)
polynomials. We arrive at the expansion $k(\rho') = \sum_{n=-\infty}^\infty g_n\ v_n(\rho')$.

The expression for the radial wave function $F+K$ becomes
\begin{eqnarray}
F+K &\equiv& \tilde{h}(\rho) e^{-\frac{1}{2}\rho} \nonumber\\
&=& \rho^{\tau-1-s} e^{-\frac{1}{2}\rho}\, \sum_{n=-\infty}^{\infty}\, g_n\, {}_1F_1(-n;1+2\tau-2s;\rho) \nonumber\\
&=& \rho^{\tau-1-s} e^{-\frac{1}{2}\rho}\, \left\{ \sum_{n=0}^{\infty}\, \frac{n!}{(1+2\tau-2s)_n} \,g_n\, L_n^{(2\tau-2s)}(\rho)
+ \sum_{n=1}^{\infty}\, g_{-n}\, {}_1F_1(n;1+2\tau-2s;\rho) \right\}
\label{eq:radialcomponent}
\end{eqnarray}
For large $\rho\rightarrow\infty$ the ${}_1F_1()$ behaves as:
\[ {}_1F_1(n;1+2\tau-2s;\rho) \sim \frac{\Gamma(1+2\tau-2s)}{(n-1)!} \rho^{n-1-2\tau+2s} e^\rho \]
So that the third boundary condition is satisfied, it is necessary that $g_n\equiv 0$
for all $n<0$.

For the eigenvalues, i.e. special values of '$y$', the CF terminates from the left,
leading to $g_{-1}=g_{-2}= ... = 0$. The eigenvalues are determined by:
\begin{eqnarray}
 0 &=& \frac{B_0}{C_0} 
 -\frac{A_1/C_1}{B_1/C_1-}\frac{A_2/C_2}{B_2/C_2-}\frac{A_3/C_3}{B_3/C_3-} .... = \frac{B_0}{C_0} + \K_{m=1}^\infty \left(\frac{-A_m/C_m}{B_m/C_m}\right)
\label{eq:eigenvalues}
\end{eqnarray}
The eigenvalues lead to global solutions, i.e. solutions that exist throughout
the entire $\rho'$-plane except for the singularities at $\rho'=1$ and
$\rho'=\infty$.\\
Note that the CF \eqref{eq:eigenvalues} is not suitable
for deriving an expansion in $\alpha^2$ for the eigenvalues. The deeper reason is that the tail
of the CF contributes to the lower terms in the expansion.
%
\subsection{Convergence of the series expansion}
\noindent The question of convergence arises for the series
\begin{equation}
\sum_{n=0}^\infty \; g_n\,v_n(\rho') = \sum_{n=0}^\infty \; g_n\,v_n(-y\rho) = \sum_{n=0}^\infty \; g_n\, {}_1F_1(-n; 1+2\tau-2s;\rho)
 = \sum_{n=0}^\infty \; g_n\, \frac{n!}{(1+2\tau-2s)_n} \,L^{(2\tau-2s)}_n(\rho)
\label{eq:series}
\end{equation}
\begin{theorem}
\label{thm:minimalsolution}
The recurrence relation for $g_n$ in equation~\eqref{eq:heunttrr} has two linearly
independent solutions $g_n^{(1)}$ and $g_n^{(2)}$ with the properties:
\begin{enumerate}[label=\alph*)]
\item
  Equation \eqref{eq:heunttrr} has one minimal solution, or
  \[ \lim_{n\to\infty} \frac{g_n^{(2)}}{g_n^{(1)}} = 0, \]
\item
  In the limit of large $n$ the solutions $g_n^{(1)}$ and $g_n^{(2)}$ behave as
  \begin{equation}
    \lim_{n\to\infty} \frac{g_{n+1}^{(1)}}{g_n^{(1)}} = 1 + \frac{1}{\sqrt{yn}} + {\cal O}(1/n), \,\,\text{and}\,
    \lim_{n\to\infty} \frac{g_{n+1}^{(2)}}{g_n^{(2)}} = 1 - \frac{1}{\sqrt{yn}} + {\cal O}(1/n)
  \end{equation}
\end{enumerate}
\begin{proof}
The result follows immediately from an extension of the Poincar\'e-Perron-Kreuser theorem formulated
in Theorem 5, subcase IIc in \cite{Koo94}.
\end{proof}
\end{theorem}
Now that we have shown that equation \eqref{eq:heunttrr} has a minimal solution, we
establish a connection between $g_n$ and the minimal solution $g_n^{(2)}$.
For completeness and convenience, we reiterate here the important theorem
that states (see Theorem 1.1 in \cite{Gau67} and Theorem 3.6.1 in \cite{Cuy10}):
\begin{theorem}[Pincherle \cite{Pin94}]
\label{thm:pincherle}
The CF \eqref{eq:ctdfraction} converges if and only if the recurrence
relation \eqref{eq:heunttrr} possesses a minimal solution $g_n^{(2)}$, with $g_0^{(2)}\neq 0$.
In case of convergence, moreover, one has
\begin{equation}
\frac{g_n^{(2)}}{g_{n-1}^{(2)}} = \frac{-A_n/C_n}{B_n/C_n+}\, \frac{-A_{n+1}/C_{n+1}}{B_{n+1}/C_{n+1}+}\, \frac{-A_{n+2}/C_{n+2}}{B_{n+2}/C_{n+2}+}\cdots,
\,\,\, n=1,2,3,\cdots,
\end{equation}
provided $g_n^{(2)}\neq 0$ for $n=0,1,2,\dots$.
\end{theorem}
In \thmref{thm:minimalsolution} we established that equation \eqref{eq:heunttrr} has a
minimal solution, and an application of \thmref{thm:pincherle} for $n=1$ then confirms
that the CF \eqref{eq:ctdfraction} converges. Furthermore, it gives the
vital result that for the minimal solution $g_n^{(2)}$:
\begin{equation}
\frac{g_1^{(2)}}{g_0^{(2)}} = \frac{-A_1/C_1}{B_1/C_1+}\, \frac{-A_2/C_2}{B_2/C_2+}\cdots,
\end{equation}
Since this is the same expression as for the ratio $g_n/g_{n-1}$ in \eqref{eq:ctdfraction},
we identify $g_n$ with the minimal solution $g_n^{(2)}$, that is $g_n\sim g_n^{(2)}$.
The eigenvalues $z=z_n$ defined by equation \eqref{eq:eigenvalues} determine this ratio and
we have
\begin{equation}
\frac{g_1}{g_0} = \frac{g_1^{(2)}}{g_0^{(2)}} = -\frac{B_0}{C_0}
\end{equation}
Finally, we are in a position to address the convergence of the series defined in \eqref{eq:series}.
To establish its convergence or divergence, we study the ratio between the $(n+1)$-term
and the $n^{\mbox{th}}$ term in the limit that $n$ tends to infinity:
\begin{equation}
    \lim_{n\rightarrow\infty} \left| \frac{g_{n+1}}{g_n}\,\frac{v_{n+1}(\rho')}{v_n(\rho')} \right| =
        1 + \frac{h}{\sqrt{n}} + \frac{B(n)}{n^r},
\label{eq:kummerstest}
\end{equation}
where $B(n)$ is bounded and $r \ge 1$. Kummer's test states that for $h < 0$ (using $a_n=\sqrt{n}$ as a
sequence of positive numbers) the series \eqref{eq:series} converges absolutely.

Next, we turn our attention to the ratio of subsequent Laguerre polynomials. For this,
the asymptotic expansion of the Laguerre polynomial is needed up to the next-to-leading term
(see \cite{Dea13}):
\begin{eqnarray}
\frac{L_{n+1}^{(\alpha)}(\rho)}{L_n^{(\alpha)}(\rho)} &=& 1 + {\cal O}(1/n) \\
\lim_{n\rightarrow\infty} \left| \frac{g_{n+1}}{g_n}\,\frac{v_{n+1}(\rho')}{v_n(\rho')} \right| &=&
        1 - \frac{1}{\sqrt{yn}} + {\cal O}(1/n)
\end{eqnarray}
From this it follows that $h=-1/\sqrt{y}$ \eqref{eq:kummerstest} and since $h<0$
the series (\ref{eq:series}) converges absolutely for all $\rho\geq0$.
%
\section{\label{sec:breitsinglet}Numerical determination of the eigenvalues}
Using the modified Lentz method \cite{Press92} for numerically evaluating the CF,
we establish over 25 digits of accuracy for the numerical values of the eigenvalues $z_n$
for the first 5 eigenvalues. A direct calculation of the CF~\eqref{eq:eigenvalues} using Lentz's method
has an abysmal performance near the eigenvalues.
Table~\ref{tab:accelerators} illustrates the slow convergence even after 2800 terms for values
very near the first eigenvalue $z_{1}$. The value of the CF should be very close
to zero. The second column shows that after the first 100 terms, hardly any improvement is made
when terms are added.

To improve upon its convergence, we have experimented with several
algorithms for speeding up the convergence of Lentz's method. In particular, from the non-linear
sequence accelerators mentioned in \cite{Wen03,Rho03} such as Peter Wynn's epsilon (including Shanks)
($\varepsilon^{(n)}$), rho ($\rho^{(n)}$), iterated rho ($\bar{\rho}^{(n)}$), Brezinski's
theta ($\vartheta^{(n)}$), and iterated theta ($\bar{\vartheta}^{(n)}$) transformations. In addition
the rho transformation ($\rho^{(n)}_\theta$) with Osada's \cite{Osa90} decay parameter $\theta$
is included. The rho transformation with a decay parameter between 20 and 32 gives the best results.
We believe that the subtractions in the recurrence relations for the iterated versions of both the rho
and theta transformations cause loss of digits and instability of these accelerators in our case.
\begin{table}[htb]
\begin{tabular}{c|c|c|c|c|c|c|c}
Terms ($n$) & {Direct (Lentz)} & $\varepsilon^{(n)}$ & $\rho^{(n)}$ & $\rho^{(n)}_{20}$ ($\theta=20$) & $\bar{\rho}^{(n)}$ & $\vartheta^{(n)}$ & $\bar{\vartheta}^{(n)}$ \\ \hline
100 & 6.48242e-08 & 2.03165e-09 &  1.34457e-11 & 3.81424e-10 & -1.86606e-11 & -2.19194e+48 & 1.08064e-08 \\
200 & 3.21444e-08 & 3.44301e-10 & -2.50072e-13 & 9.11380e-12 & 2.31295e-12 & 1.59197e+257 & -1.45011e-10 \\
300 & 2.12086e-08 & 9.43581e-11 & -7.96823e-14 & 3.20667e-13 & 4.41809e-12 & 7.34845e+371 & -1.10188e-10 \\
400 & 1.57417e-08 & 3.12217e-11 &  1.17139e-14 & 8.36056e-15 & 4.36732e-12 & -6.79198e+478 & -1.10270e-10 \\
500 & 1.24659e-08 & 1.14852e-11 & -5.77508e-16 & 1.62133e-16 & 4.39731e-12 & -4.57233e+573 & -1.10270e-10 \\
1000 & 5.94714e-09 & 1.74122e-13 & -7.76667e-20 & -2.51454e-22 & 4.25503e-12 & 4.06630e+1075 & 1.77288e-12 \\
1500 & 3.80118e-09 & 5.33004e-15 & -9.11731e-24 & 5.30267e-27 & 4.17799e-12 & 1.77141e+1478 & 1.36991e-12 \\
2000 & 2.74201e-09 & 2.38678e-16 &  7.06438e-27 & 5.30682e-27 & 4.04126e-12 & 1.09739e+1793 & 1.37121e-12 \\
2500 & 2.11475e-09 & 1.37636e-17 &  5.30869e-27 & 5.30682e-27 & 4.00509e-12 & -9.51281e+2027 & -2.49691e-14 \\
2800 & 1.85432e-09 & 2.72460e-18 &  5.30689e-27 & 5.30682e-27 & 3.98258e-12 & -1.69965e+1907 & -2.33865e-14
\end{tabular}
\caption{Comparison between several sequence accelerators for the calculation of the CF near the
value $z=0.9999999968218806697141710$ close to the first eigenvalue.\label{tab:accelerators}}
\end{table}
Furthermore, to overcome the suffering of the loss of digits in the rho transformation, the calculations
need to be carried out at high precision (over 6000 digits) to calculate the eigenvalues for the
first five energy levels to obtain 25 digits of accuracy. Based on the results in table \ref{tab:accelerators}
we have chosen the $\rho$-algorithm to calculate values of the CF \eqref{eq:eigenvalues}.\\

Solutions to \eqref{eq:eigenvalues} give the eigenvalues and the energy levels. Figure~\ref{fig:ctdfraction}
shows a plot of the CF for various values of $z$. As can be inferred from the
figure, the values of $z$ for which the CF vanishes are all near the points
$n=1, 2, 3, ....$. Figure~\ref{fig:cfracdetails} shows the structure~\eqref{eq:eigenvalues} near its zero
and asymptotic. Starting from $z=4$, the same pattern repeats itself for higher $z$ as well. Between
$0<z<3$, the CF behaves quite regularly and is numerically stable to calculate. Around $z=3$, the
behaviour is most complex, and calculations need to be carried out with high precision.\\
\begin{figure}[htb]
\begin{center}
\includegraphics[height=2in,width=4in,angle=0]{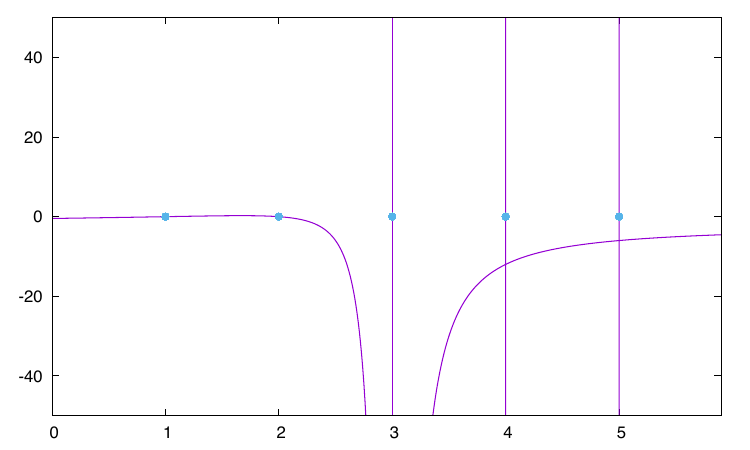}
\caption{Continued fraction for the range $z=0...6$. The dots mark the location of the eigenvalues.\label{fig:ctdfraction}}
\end{center}
\end{figure}
\begin{figure}[htb]
\begin{center}
\begin{subfigure}{0.2\textwidth}
\includegraphics[width=\textwidth]{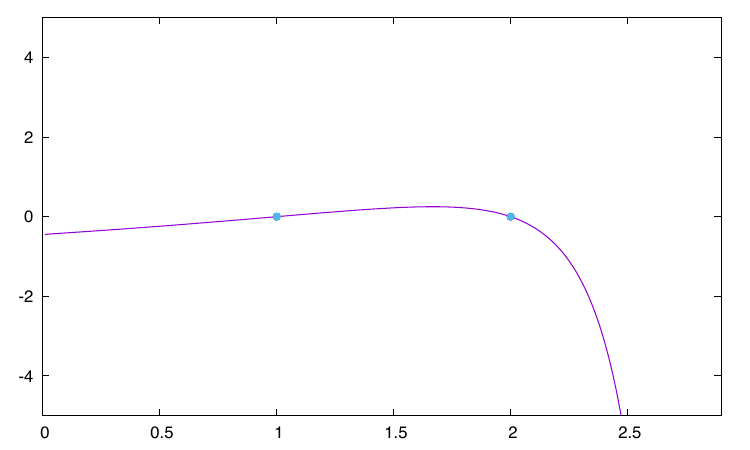}
\caption{Values of~\eqref{eq:ctdfraction} for $0<z<2.9$.}
\end{subfigure}
\begin{subfigure}{0.2\textwidth}
\includegraphics[width=\textwidth]{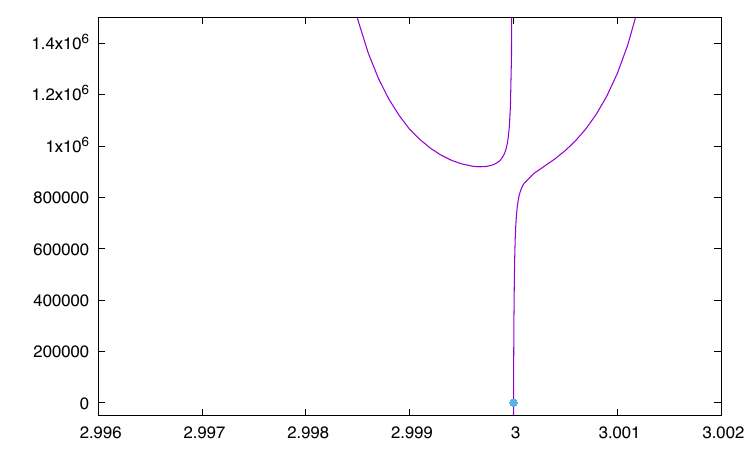}
\caption{Values of~\eqref{eq:ctdfraction} for $2.9<z<3.1$.}
\end{subfigure}
\begin{subfigure}{0.2\textwidth}
\includegraphics[width=\textwidth]{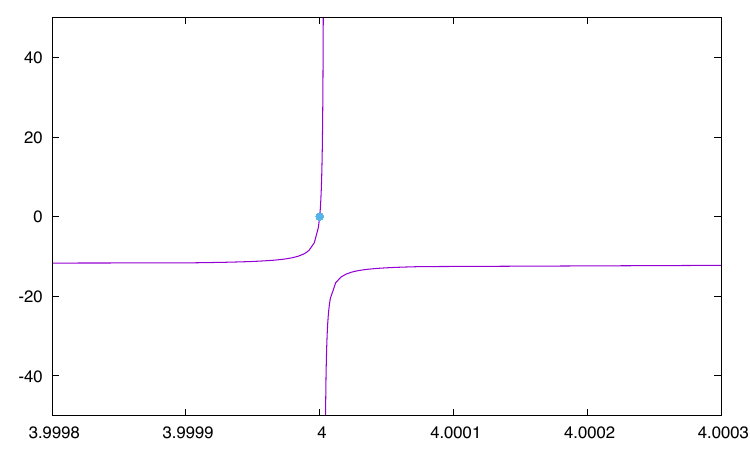}
\caption{Values of~\eqref{eq:ctdfraction} for $3.9<z<4.1$.}
\end{subfigure}
\begin{subfigure}{0.2\textwidth}
\includegraphics[width=\textwidth]{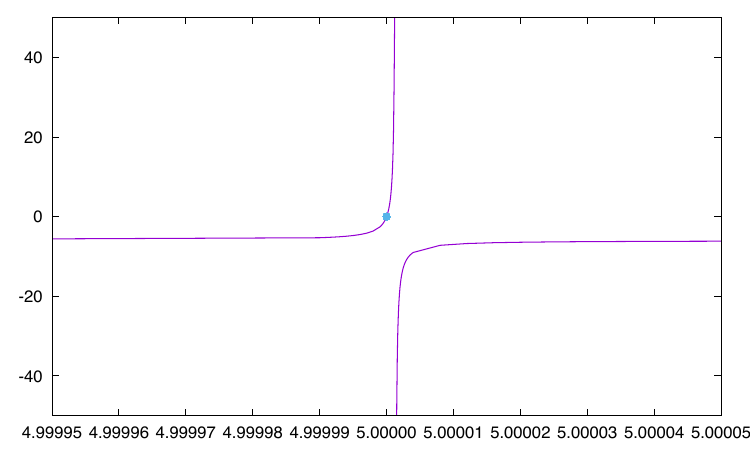}
\caption{Values of~\eqref{eq:ctdfraction} for $4.9<z<5.1$.}
\end{subfigure}
\end{center}
\caption{Detailed graphs of Figure~\ref{fig:ctdfraction} around the zeros and asymptotic behaviour of~\eqref{eq:eigenvalues}.
The dots mark the location of the eigenvalues.}
\label{fig:cfracdetails}
\end{figure}
From~\eqref{eq:eigenvalues}, the eigenvalues have been determined from their zeros. The sequence of
partial products from the modified Lentz method has been accelerated using Peter Wynn's non-linear
$\rho$ transformation \cite{Wen03} and combined with the secant method for root-finding. The
calculations have been carried out with 20000 bits of precision using the Python package {\tt gmpy2} \cite{gmpy2}.
The results are shown in Table~\ref{tab:energylevels}.\\

Here $E_n$ is given by \eqref{eq:energylevels}. The first 3 values coincide with the results in \cite{Sco92}
up to 18 digits. Here, we have compared our results to results obtained earlier in the literature. Note
that the work of Scott {\it et. al.}~\cite{Sco92} uses an $\alpha^{-1}=137$ while Yamaguchi \etal\ 
in \cite{Yam01} uses a value of $\alpha^{-1}=137.035999765$ for which we have corrected in
Table~\ref{tab:energylevels}.\\

In table \ref{tab:eigenlevels} we compare our results for the eigenvalues directly with those of Yamaguchi
\etal~\cite{Yam01}. In the ordering of the energy levels, we know that the levels for the singlet states
$2{}^{1}P_{1}$ and $2{}^{1}S_{0}$ are reversed \cite{Adk22}. The reason is that equation \eqref{eq:1.1}
does not include the Breit interaction. The spacing of the energy levels
$2{}^1S_{0} - 1{}^1S_{0}=0.000004994876$ which is close to the value $0.000004991256$ derived from
table 2 in \cite{Adk22}.
\begin{table}[htb]
\begin{tabular}{c|c|c|c|c|c|c|c}
$n$ & $\ell$ & $s$ & $j$ & $z_n \text{(this work)}$ & $E_n \text{(this work)}$ & $E_n \text{(FEM~\cite{Sco92})}$ & $E_n$ \text{(Tursunov \etal \cite{Tur24})} \\ \hline
1 & 0 & 0 & 0 &  $0.9999999968218806697141710$ & $0.999993340148538880\ 1217699$ & $0.999993340148538880$          & $0.999 993 343 356 843$ \\ \hline
2 & 0 & 0 & 0 &  $1.9999999967755540467085545$ & $0.999998335024665402\ 1773807$ & $0.999998335024665402$          & \\
2 & 1 & 0 & 1 &  $1.9999955600612685817892653$ & $0.999998335017278376\ 1818110$ & $0.999998335017278391$          & $0.999 998 342 513 543$ \\ \hline
3 & 0 & 0 & 0 &  $2.9999999967672139077910220$ & $0.999999260009936471\ 6125118$ & $0.999999260009936472$          & \\
3 & 1 & 0 & 1 &  $2.9999955600666057950709621$ & $0.999999260007747726\ 2743266$ & $0.999999260007747730$          & \\
3 & 2 & 0 & 2 &  $2.9999973360318569473896816$ & $0.999999260008623859\ 5046031$ & $0.999999260008623860$          & \\ \hline
4 & 0 & 0 & 0 &  $3.9999999967643175837252131$ & $0.999999583755387352\ 9138206$ &                                 & \\
4 & 1 & 0 & 1 &  $3.9999955600698868289150709$ & $0.999999583754463976\ 8759419$ &                                 & \\
4 & 2 & 0 & 2 &  $3.9999973360320343645074989$ & $0.999999583754833594\ 9673024$ &                                 & \\
4 & 3 & 0 & 3 &  $3.9999980971658464620562330$ & $0.999999583754992004\ 0638187$ &                                 & \\ \hline
5 & 0 & 0 & 0 &  $4.9999999967629813105957135$ & $0.999999733603388113\ 4032523$ &                                 & \\
5 & 1 & 0 & 1 &  $4.9999955600721069793318352$ & $0.999999733602915345\ 1958403$ &                                 & \\
5 & 2 & 0 & 2 &  $4.9999973360321655244252866$ & $0.999999733603104589\ 4202622$ &                                 & \\
5 & 3 & 0 & 3 &  $4.9999980971658680971638007$ & $0.999999733603185694\ 8746888$ &                                 & \\
5 & 4 & 0 & 4 &  $4.9999985200180055246694377$ & $0.999999733603230753\ 4530486$ &                                 & \\
\end{tabular}
\caption{Approximations of the first five energy levels in units of the electron mass $m_e$ and
eigenvalues. The estimated error is $10^{-25}$.\label{tab:energylevels}}
\end{table}
\begin{table}[htb]
\begin{tabular}{c|c|c|c|c|c|c|c}
$n$ & $\ell$ & $s$ & $j$ & $z_n \text{(this work)}$ & $z_n \text{(Yamaguchi \etal \cite{Yam01})}$ \\ \hline
1 & 0 & 0 & 0 & $0.9999999968218806697141710$ & $0.99999999599459699793$ \\ \hline
2 & 0 & 0 & 0 & $1.9999999967755540467085545$ & $2.0000000345605969979$  \\
\end{tabular}
\caption{Approximations of the first five energy levels in units of the electron mass $m_e$ and
eigenvalues. The estimated error is $10^{-25}$.\label{tab:eigenlevels}}
\end{table}
\subsection{Normalization and calculation of the eigenfunctions}
We normalize the eigenfunctions \eqref{eq:series} by exploiting the boundary condition
\eqref{eq:boundarycond2}. This leads to the condition on the $g_n$, and using the orthogonality
properties of the Laguerre polynomials
\begin{equation}
\int_0^\infty \rho^{2\tau-2s}\, e^{-\rho}\, \left( L_n^{(2\tau-2s)}(\rho) \right)^2 d\rho = \frac{\Gamma(n+2\tau-2s+1)}{n!}
\end{equation}
we find the condition on $g_n$:
\begin{equation}
\int_0^\infty \rho^2\, \left|F(\rho)+K(\rho)\right|^2 d\rho = \Gamma(2\tau-2s+1) \sum_{n=0}^\infty g_n^2\, \frac{n!}{(1+2\tau-2s)_n} = \left(\frac{2}{z}\frac{1}{2a_0}\frac{E}{2m}\right)^3
\label{eq:conditiongn}
\end{equation}

With the results obtained in the previous section, it is possible to calculate the values of the
radial components based on the position $r$ and equations \eqref{eq:radialcomponent}
and \eqref{eq:series}. An efficient algorithm is Miller's backward recurrence algorithm
for the non-linear condition \eqref{eq:conditiongn}. For details on the algorithm, see
(3.9) and (3.9p) in \cite{Gau67}.

\begin{figure}[htb]
\begin{center}
\includegraphics[width=\textwidth]{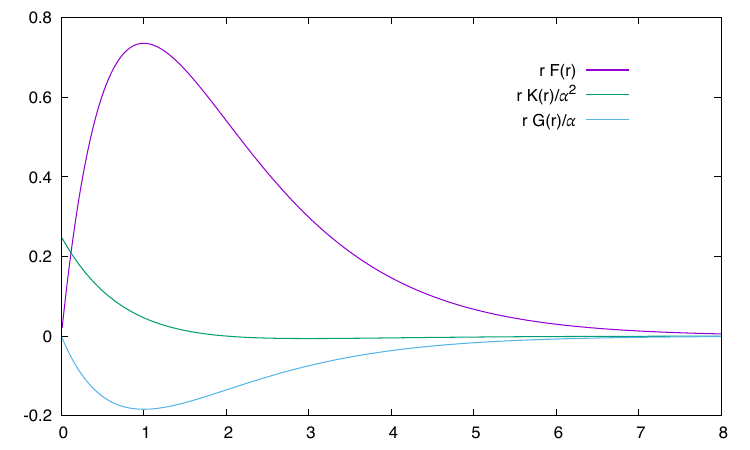}
\caption{The scaled radial components $F(r)$, $K(r)$, and $G(r)$ for the spin singlet case $N=1$. The position $r$
is expressed in units $2a_0$.\label{fig:radialn1l0s0j0_FKG}}
\end{center}
\end{figure}

Figure \ref{fig:radialn1l0s0j0_FKG} depicts the scaled radial component large-large $F(r)$ \eqref{eq:1.7} in units
of the Bohr radius $2a_0$. The large-small component $K(r)$, and the small-small component $G(r)$.
Compare these to the solid and dashed curves in FIG 1 \cite{Sco92}.

As discussed in FIG 3 \cite{Sco92}, the scaled small-small radial component $rK(r)$ goes to zero
at $r=0$ despite that this is not apparent from Figure \ref{fig:radialn1l0s0j0_FKG}. The reason is
that it approaches its maximum around 1 fermi and drops to zero starting at $\rho\sim 1/y$.\\
Only recently has this been discussed by Patterson in \cite{Pat19}. Patterson shows that this
behaviour of the wave function near the point $\rho=1/y$ is not present when calculations are
based on the relativistic Bethe-Salpeter equation and therefore corresponds to a spurious state.
%

\section{\label{sec:spintriplet}The spin-triplet states}
Next, we discuss the spin-triplet states and the numerical results. In this section, we
repeat the analysis for the radial functions in \cite{Sco92} and \cite{Mal88} and
show that the spin-singlet state and all the spin-triplet states, except for states with
$J=L\pm1$, can be described by a set of three variables. The importance of this is that
the methods of solving for the eigenvalues for the spin-single state can be extended to
solving for most of the spin-triplet states. In particular, this applies to the methods
based on the CF \eqref{eq:eigenvalues} and determinant equation
\eqref{eq:deteqndelta} presented in the following sections.
%
\subsection{\label{sec:tripletjiszero}The spin-triplet $J=0$ case (${}^3P_0$)}
\begin{figure}[htb]
\begin{center}
\includegraphics[width=\textwidth]{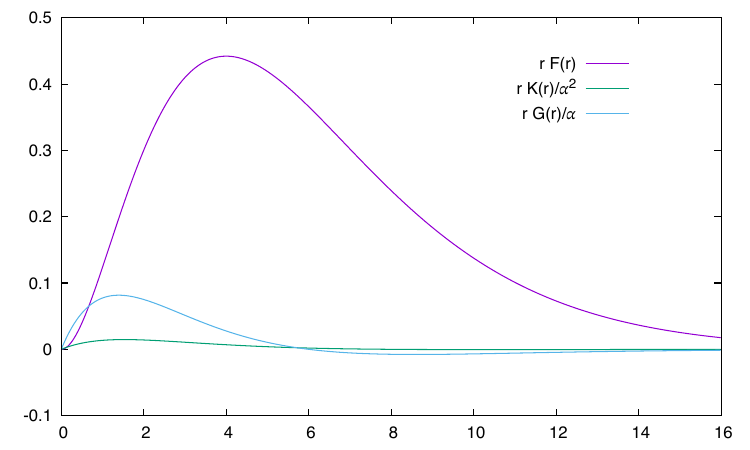}
\caption{The scaled radial components $F(r)$, $K(r)$, and $G(r)$ for the spin triplet case $N=2$. The position $r$
is expressed in units $2a_0$.\label{fig:radialn2l1s1j0_FKG}}
\end{center}
\end{figure}

In the spin-triplet case with $J=0$ the $\Psi_{\alpha\beta}$ can be taken as:
\begin{equation}
\begin{tabular}{cc|c}
$\alpha$ & $\beta$ & $\Psi_{\alpha\beta}$ \\ \hline
1 & 1 & \multirow{2}*{$F_2(r)\ket{{}^3P_0}$} \\
2 & 2 & \\ \hline
1 & 3 & \multirow{4}*{$G_2(r)\ket{{}^1(L)_L}$} \\
2 & 4 & \\ \cline{1-2}
3 & 1 & \\
4 & 2 & \\ \hline
3 & 3 & \multirow{2}*{$K_2(r)\ket{{}^3P_0}$} \\
4 & 4 & 
\end{tabular}
\end{equation}
The function $G_2(r)$ is fixed by the functions $F_2(r)$ and $K_2(r)$.
\begin{eqnarray}
&& \left(E+\frac{\alpha}{r}+2m\right) F_{2}(r) = \left(E+\frac{\alpha}{r}-2m\right) K_{2}(r) \\
&& -\left(E+\frac{\alpha}{r}\right) G_{2}(r) = \left( \frac{d}{dr} + \frac{j+2}{r} \right) (F_2(r)+K_2(r))
\end{eqnarray}
Changing variables to $\rho$ \eqref{eq:1.4a} and defining the function $\tilde{h}_2(\rho)$ by
\[ F_{2}(\rho) + K_{2}(\rho) = \tilde{h}_{2}(\rho) \exp\left(-\frac{1}{2}\rho\right) \]
the differential equation satisfied by $\tilde{h}_{2}(\rho)$ is
\begin{equation}
\left\{ \frac{d^2}{d\rho^2} + \left( -1 + \frac{2}{\rho} + \frac{1}{\rho(y\rho+1)} \right)\frac{d}{d\rho} + \frac{-\frac{1}{2}}{\rho(y\rho+1)} - \frac{2y}{\rho(y\rho+1)} + \frac{\frac{1}{2}\alpha^2 y - 1}{\rho} + \frac{\frac{1}{4}\alpha^2}{\rho^2} \right\} \tilde{h}_2(\rho) = 0
\end{equation}
Near the origin $\rho=0$ the radial component behaves as $\rho^{\tau-1-s}$ with
$\tau=1$ and $s$ defined as $s=1-\sqrt{1 - \frac{1}{4}\alpha^2}$.
\begin{equation}
\left\{ \frac{d^2}{d\rho^2} + \left( -1 + \frac{3 - 2s}{\rho} - \frac{y}{y\rho+1} \right)\frac{d}{d\rho} + \frac{y\rho(\frac{1}{2}\alpha^2 y - 1 + s) + (s-2)(1+y) + \frac{1}{2} + \frac{1}{2}\alpha^2 y}{\rho(1+y\rho)} \right\} k_2(\rho) = 0
\end{equation}
where $k(\rho)$ is defined as $\tilde{h}_2(\rho) = \rho^{\tau-1-s}\,k_2(\rho)$. From the last
term we recognize $\sigma$ and $\tau$ (see also \eqref{eq:sigmasinglet}):
\begin{subequations}
\begin{eqnarray}
&&\sigma \rightarrow 2-s + \left( -\frac{1}{2}\alpha^2 y + \frac{1}{2} + \tau - s \right)/y \\
&&\tau \rightarrow 1
\end{eqnarray}
\label{eq:sigmatripletjiszero}
\end{subequations}
Table~\ref{tab:energylevelstripletjiszero} compares our results to results obtained earlier
in the literature \cite{Sco92,Tur24}. The radial components large-large ($rF(r)$), large-small ($rG(r)$),
and small-small ($rK(r)$) are displayed in Figure \ref{fig:radialn2l1s1j0_FKG}.
\begin{table}[htb]
\begin{tabular}{c|c|c|c|c|c|c|c}
$n$ & $\ell$ & $s$ & $j$ & $z_n \text{(this work)}$      & $E_n \text{(this work)}$        & $E_n \text{(FEM~\cite{Sco92})}$ & $E_n$ \text{(Tursunov \etal \cite{Tur24})} \\ \hline
2   & 1      & 1   & 0   & $1.9999911200757235154307452$ & $0.999998335009885854\ 3994827$ & $0.999998335009885854$ & $0.999 998 342 584 549$ \\ \hline
3   & 1      & 1   & 0   & $2.9999911200798304738000166$ & $0.999999260005557350\ 0586677$ & $0.999999260005557351$ & \\ \hline
4   & 1      & 1   & 0   & $3.9999911200812679043058628$ & $0.9999995837535399121673290$ & & \\ \hline
5   & 1      & 1   & 0   & $4.9999911200819332280824889$ & $0.9999997336024422241593037$ & &
\end{tabular}
\caption{Approximations of the first three energy levels in units of the electron mass $m_e$ and
eigenvalues for spin-triplet and $J=0$. The estimated error is $10^{-25}$.\label{tab:energylevelstripletjiszero}}
\end{table}
%
\subsection{\label{sec:tripletjisl}The spin-triplet $J=L$ case (${}^3J_J$)}
\begin{figure}[htb]
\begin{center}
\includegraphics[width=\textwidth]{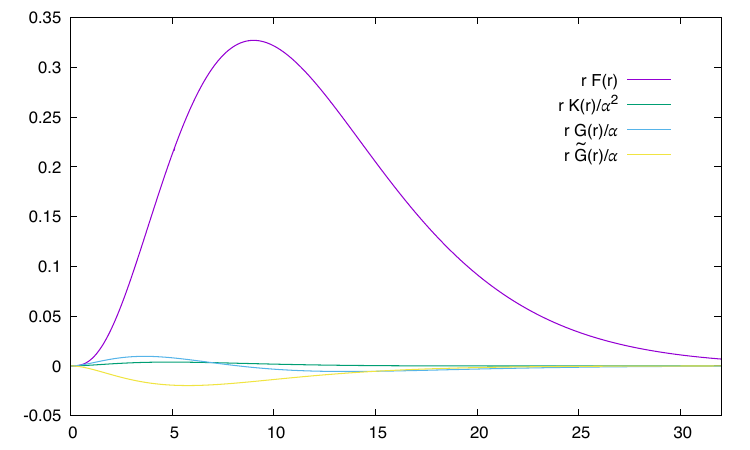}
\caption{The scaled radial components $F(r)$, $K(r)$, $G(r)$, and $\tilde{G}$ for the spin triplet case $N=3, L=2, S=1, J=2$.
The position $r$ is expressed in units $2a_0$.\label{fig:radialn3l2s1j2_FKG}}
\end{center}
\end{figure}

The radial component functions $\hat{F}(r)$ and $\hat{K}(r)$ are defined by
combinations of $\Psi_{\alpha\beta}$:
\begin{equation}
\begin{tabular}{cc|c}
$\alpha$ & $\beta$ & $\Psi_{\alpha\beta}$ \\ \hline
1 & 1 & \multirow{2}*{$F_3(r)\ket{{}^3(L)_L}$} \\
2 & 2 & \\ \hline
1 & 3 & \multirow{4}*{$i\left\{ G_3(r)\ket{{}^3(L+1)_L} + \tilde{G}_3(r)\ket{{}^3(L-1)_L} \right\}$} \\
2 & 4 & \\ \cline{1-2}
3 & 1 & \\
4 & 2 & \\ \hline
3 & 3 & \multirow{2}*{$K_3(r)\ket{{}^3(L)_L}$} \\
4 & 4 & 
\end{tabular}
\end{equation}
The functions $F_3(r)$ and $K_3(r)$ completely fix the functions $G_3(r)$
and $\tilde{G}_3(r)$:
\begin{eqnarray}
\left(E+\frac{\alpha}{r}-2m\right) F_{3}(r) &=& \left(E+\frac{\alpha}{r}+2m\right) K_{3}(r) \\
\left(E+\frac{\alpha}{r}\right) (G_{3}(r) + \tilde{G}_{3}(r)) &=& \sqrt{\frac{j}{2j+1}} \left( \frac{d}{dr} - \frac{j}{r} \right) (F_3(r)+K_3(r)) \\
\left(E+\frac{\alpha}{r}\right) (G_{3}(r) - \tilde{G}_{3}(r)) &=& \sqrt{\frac{j+1}{2j+1}} \left( \frac{d}{dr} + \frac{j+1}{r} \right) (F_3(r)+K_3(r))
\end{eqnarray}
Changing variables to $\rho$ \eqref{eq:1.4a} and defining the function $\tilde{h}_{3}(\rho)$ by
\[ F_{3}(\rho) + K_{3}(\rho) = \tilde{h}_{3}(\rho) \exp\left(-\frac{1}{2}\rho\right) \]
the differential equation satisfied by $\tilde{h}_3(\rho)$ is
\begin{equation}
\left\{ \frac{d^2}{d\rho^2} + \left( -1 + \frac{2}{\rho} + \frac{1}{\rho (y\rho+1)}\right)\frac{d}{d\rho} + \frac{-\frac{1}{2}}{\rho (y\rho+1)} +  \frac{1}{\rho^2 (y\rho+1)} + \frac{\frac{1}{2}\alpha^2 y - 1}{\rho} + \frac{\frac{1}{4}\alpha^2  - j(j+1)}{\rho^2} \right\} \tilde{h}_3(\rho) = 0
\end{equation}
Near the origin $\rho=0$ the radial component behaves as $\rho^{\tau-1-s}$ with
$\tau=\sqrt{j(j+1)}$ and $s$ defined as $s=\tau-\sqrt{\tau^2 - \frac{1}{4}\alpha^2}$.
\begin{equation}
\left\{ \frac{d^2}{d\rho^2} + \left( -1 + \frac{3+2\sigma}{\rho} - \frac{y}{y\rho+1}\right)\frac{d}{d\rho} + \frac{y\rho(\frac{1}{2}\alpha^2 y - \tau + s) + \frac{1}{2}\alpha^2 y - \frac{1}{2} - (\tau-s) (y+1)}{\rho(1+y\rho)} \right\} k_3(\rho) = 0
\end{equation}
where $k_3(\rho)$ is defined as $\tilde{h}_3(\rho) = \rho^{\tau-1-s}\,k_3(\rho)$. From the last
term we recognize $\sigma$ and $\tau$ (see also \eqref{eq:sigmasinglet}):
\begin{subequations}
\begin{eqnarray}
&&\sigma \rightarrow  \tau-s + \left( -\frac{1}{2}\alpha^2 y + \frac{1}{2} + \tau - s \right)/y \\
&&\tau \rightarrow \sqrt{j(j+1)}
\end{eqnarray}
\label{eq:sigmatripletjisl}
\end{subequations}
Table~\ref{tab:energylevelstripletjisl} compares our results to results obtained earlier
in the literature \cite{Sco92}. Graphs of the four radial components $rF(r)$ (large-large), $rK(r)$ (small-small),
$rG(r)$ (large-small), and $r\tilde{G}(r)$ (large-small) are shown in Figure \ref{fig:radialn3l2s1j2_FKG}.
\begin{table}[htb]
\begin{tabular}{c|c|c|c|c|c|c}
$n$ & $\ell$ & $s$ & $j$ & $z_n \text{(this work)}$      & $E_n \text{(this work)}$        & $E_n \text{(FEM~\cite{Sco92})}$ \\ \hline
2   & 1      & 1   & 1   & $1.9999933400894389935998919$ & $0.999998335013582156\ 3146492$ & $0.999998335013582156$ \\ \hline
3   & 1      & 1   & 1   & $2.9999933400939558923891662$ & $0.999999260006652549\ 6126998$ & $0.999999260006652550$ \\
3   & 2      & 1   & 2   & $2.9999968920374259933553312$ & $0.999999260008404824\ 8046296$ & $0.999999260008404825$ \\ \hline
4   & 1      & 1   & 1   & $3.9999933400955368033219382$ & $0.9999995837540019490602989$ & \\
4   & 2      & 1   & 2   & $3.9999968920376929405687525$ & $0.9999995837547411896988064$ & \\
4   & 3      & 1   & 3   & $3.9999979385963910799081262$ & $0.9999995837549590021946505$ & \\ \hline
5   & 1      & 1   & 1   & $4.9999933400962685383242704$ & $0.9999997336026787868862155$ & \\
5   & 2      & 1   & 2   & $4.9999968920378164989152288$ & $0.9999997336030572779211945$ & \\
5   & 3      & 1   & 3   & $4.9999979385964338502455569$ & $0.9999997336031687979173547$ & \\
5   & 4      & 1   & 4   & $4.9999984460189226652513691$ & $0.9999997336032228682070500$ & \\
\end{tabular}
\caption{Approximations of the first three energy levels in units of the electron mass $m_e$ and
eigenvalues for spin-triplet and $J=L$. The estimated error is $10^{-25}$.\label{tab:energylevelstripletjisl}}
\end{table}
%
\subsection{\label{sec:tripletjislpmone}The spin-triplet $J=L\pm1$ case (${}^3(L\pm 1)_{L}$)}
For completeness, we present the differential equations for the radial components for the
spin-triplet state and $J=L\pm 1$. The radial component functions $\bar{G}(r)$, and $\bar{H}(r)$
are defined by combinations of $\Psi_{\alpha\beta}$:
\begin{equation}
\begin{tabular}{cc|c}
$\alpha$ & $\beta$ & $\Psi_{\alpha\beta}$ \\ \hline
1 & 1 & \multirow{2}*{$F_2(r)\ket{{}^3(L+1)_L} + \tilde{F}_2(r)\ket{{}^3(L-1)_L}$} \\
2 & 2 & \\ \hline
1 & 3 & \multirow{4}*{$i\,G_2(r)\ket{{}^1(L)_L}$,\,\, $i\,\tilde{G}_2(r)\ket{{}^3(L)_L}$} \\
2 & 4 & \\ \cline{1-2}
3 & 1 & \\
4 & 2 & \\ \hline
3 & 3 & \multirow{2}*{$K_2(r)\ket{{}^3(L+1)_L} + \tilde{K}_2(r)\ket{{}^3(L-1)_L}$} \\
4 & 4 & 
\end{tabular}
\end{equation}
Once the radial components $G_2(r)$ and $\tilde{G}_2(r)$ are known the other functions are
determined by:
\begin{eqnarray}
\left(E+\frac{\alpha}{r}-2m\right) F_2(r) &=& 2\sqrt{\frac{j+1}{2j+1}} \left( \frac{d}{dr} - \frac{j}{r} \right) G_2(r) + 2\sqrt{\frac{j}{2j+1}} \left( \frac{d}{dr} - \frac{j}{r} \right) \tilde{G}_2(r) \\
\left(E+\frac{\alpha}{r}+2m\right) K_2(r) &=& -2\sqrt{\frac{j}{2j+1}} \left( \frac{d}{dr} + \frac{j+1}{r} \right) G_2(r) + 2\sqrt{\frac{j+1}{2j+1}} \left( \frac{d}{dr} + \frac{j+1}{r} \right) \tilde{G}_2(r) \\
\left(E+\frac{\alpha}{r}+2m\right) \tilde{K}_2(r) &=& -4\sqrt{\frac{j+1}{2j+1}}\left( \frac{d}{dr} - \frac{j}{r} \right) G_2(r)\\
\left(E+\frac{\alpha}{r}-2m\right) \tilde{F}_2(r) &=&  4\sqrt{\frac{j}{2j+1}}\left( \frac{d}{dr} + \frac{j+1}{r} \right) G_2(r)
\end{eqnarray}
In the spin-triplet state with mixing of the states with $j=\ell\pm1$, equation \eqref{eq:1.1}
leads to two coupled differential equations for the radial components $\hat{h}_1(\rho)$
and $\hat{h}_2(\rho)$:
\begin{subequations}
\begin{multline}
\left\{ \frac{d^2}{d\rho^2} + \left( -1 + \frac{2}{\rho} \right. - \frac{1}{\rho(1+y\rho)}\right) \frac{d}{d\rho} + \frac{\frac{1}{2}\alpha^2 y - 1}{\rho} + \frac{\frac{1}{4}\alpha^2 - j(j+1)}{\rho^2}\\
\left. + \frac{\frac{1}{2}}{\rho(1+y\rho)} - \frac{\alpha^4 (\eta_{+}+\eta_{-})^2}{(-\eta_{-}\rho+\alpha^2)^2(\eta_{+}\rho+\alpha^2)^2} \right\} \hat{h}_1(\rho) - \frac{\sqrt{j(j+1)}}{\rho} \frac{\alpha^2 (\eta_{+}+\eta_{-})}{(-\eta_{-}\rho+\alpha^2)(\eta_{+}\rho+\alpha^2)} \hat{h}_2(\rho) = 0
\end{multline}
\begin{multline}
- \frac{\sqrt{j(j+1)}}{\rho} \frac{\alpha^2 (\eta_{+}+\eta_{-})}{(-\eta_{-}\rho+\alpha^2)(\eta_{+}\rho+\alpha^2)} \hat{h}_1(\rho) + \left\{ \frac{d^2}{d\rho^2} + \left( -1 + \frac{2}{\rho} - \frac{1}{\rho(1+y\rho)}\right) \frac{d}{d\rho} + \frac{\frac{1}{2}\alpha^2 y - 1}{\rho} + \frac{\frac{1}{4}\alpha^2-j(j+1)}{\rho^2} \right.\\
\left. + \frac{\frac{1}{2}}{\rho(1+y\rho)} - \frac{\alpha^4 (\eta_{+}+\eta_{-})^2}{(-\eta_{-}\rho+\alpha^2)^2(\eta_{+}\rho+\alpha^2)^2} + \frac{1}{\rho^2(1+y\rho)}\frac{(2\alpha^4 y^2+\alpha^2)\rho^2 + 2\alpha^4 y\rho + \alpha^4}{(-\eta_{-}\rho+\alpha^2)(\eta_{+}\rho+\alpha^2)}\right\} \hat{h}_2(\rho) = 0
\end{multline}
\end{subequations}
and the $\eta$'s defined as:
\[ \eta_+ = \sqrt{\alpha^2+\alpha^4 y^2} + \alpha^2 y \sim {\cal O}(1), \,\, \eta_- = \sqrt{\alpha^2+\alpha^4 y^2} - \alpha^2 y \sim {\cal O}(\alpha^2) \]
Since $\eta_{-}>0$ an additional regular singularity appears at $\rho=\eta_{+}$.
Furthermore, these equations decouple for $J=0$ (${}^3P_0$), which has been discussed in section \ref{sec:tripletjiszero}.

Near the origin $\rho=0$ and expanding $\tilde{h}_1\sim\rho^{\sigma_1}$
and $\tilde{h}_2\sim\rho^{\sigma_2}$, two solutions exist.\\

Solution 1: $\sigma_2=\sigma_1-1$
\[ \sigma_1 = 1 + \sqrt{j(j+1) - 1 - \frac{1}{4}\alpha^2}, \,\, \sigma_1 = \sqrt{j(j+1) - 1 - \frac{1}{4}\alpha^2} \]
Solution 2: $\sigma_1=\sigma_2-1$
\[ \sigma_1 = \sqrt{j(j+1) - \frac{1}{4}\alpha^2}, \,\, \sigma_2 = 1 + \sqrt{j(j+1) - \frac{1}{4}\alpha^2} \]
%
\subsection{\label{sec:sigmataumodel}Modeling of the parameters of the differential equation for $k(\rho)$}
For the various spin states equations \eqref{eq:sigmasinglet},
\eqref{eq:sigmatripletjiszero}, and \eqref{eq:sigmatripletjisl} can be summarized
in one equation by:
\begin{subequations}
\begin{eqnarray}
&&\sigma \rightarrow \tau-1 - s - \lambda + \frac{1}{2y} - \left( \frac{1}{2}\alpha^2y - \tau + s \right)/y \\
&&\tau \rightarrow \sqrt{1+j(j+1)-\kappa}
\end{eqnarray}
\label{eq:sigmaoneequation}
\end{subequations}
Here, the parameters are defined by:
\begin{eqnarray}
&&s = \tau - \sqrt{\tau^2 - \frac{1}{4}\alpha^2} \nonumber\\
&&\lambda = \ldots = \frac{1}{2}\left( J(J+1) - L(L+1) - S(S+1) \right) \nonumber\\
&&\kappa = \begin{cases}
0 & \text{for spin-singlet and spin-triplet ${}^3P_0$} \\
1 & \text{for spin-triplet $J=L$, and $J=L\pm1$ (solution 2)} \\
2 & \text{for spin-triplet $J=L\pm1$ (solution 1)} \\
\end{cases}
\label{eq:sigmaparameters}
\end{eqnarray}

\section{\label{sec:green}An integral equation based on the Green function method}
In the previous section, we established a CF \eqref{eq:eigenvalues} starting from
the differential equation \eqref{eq:2.9}. The CF has been solved numerically to
obtain approximations of the eigenvalues for the first five energy levels (see
Table~\ref{tab:energylevels}).

We discovered an alternative approach for studying the eigenvalues. The
outline of the procedure is as follows. In this section, we derive an integral
equation \eqref{eq:integraleqn} that is equivalent to the differential equation \eqref{eq:basicform}.
In section~\ref{sec:determinant} we expand the solution to the integral equation in terms
of associated Laguerre polynomials, leading to a condition for the eigenvalues expressed in
a semi-infinite determinant. Section~\ref{sec:dirac} compares the Breit and Dirac equations
and shows how the eigenvalues for the Dirac case follow from the integral
equation \eqref{eq:integraleqn}. Next, section~\ref{sec:detconv} establishes the convergence
of the semi-infinite determinant. Finally, in section~\ref{sec:smally} we show that for the
spin-singlet case the eigenvalues can be calculated to high accuracy.
\subsection{Introduction}
Consider the differential equation\footnote{Note that the $\alpha$ in \eqref{eq:inhomeqn}
is not the fine-structure constant but the parameter of the associated Laguerre polynomials
$L_n^{(\alpha)}(\rho)$. It is customary to use the letter $\alpha$ for this parameter,
e.g. see \cite{AS70}. In this and the following sections $\alpha$ indicates the parameter
of the associated Laguerre polynomials. The fine-structure constant is implicitly present
through the variables $s$ \eqref{eq:staudefinition} and $y$ \eqref{eq:zdefinition}.
Expansions in either $s$ or $1/y$ are in fact expansions in the fine-structure constant.}
\begin{equation}
\rho\frac{d^2}{d\rho^2}f(\rho) + (\alpha+1 - \rho)\frac{d}{d\rho}f(\rho) - a f(\rho) = F(\rho)
\label{eq:inhomeqn}
\end{equation}
where $F(\rho)$ is some function and the constant $a$ can take on any value except for eigenvalues
of the differential operator.

The solution to this equation, expressed using the Green function, is:
\begin{equation}
f(\rho) = \int_0^\infty d\rho' G(\rho,\rho') F(\rho')
\label{eq:greensolution}
\end{equation}
An explicit expression for the Green function $G(\rho,\rho')$ is
\[ G(\rho,\rho') = -\sum_{n=0}^\infty \frac{n!}{\Gamma(\alpha+1+n)} \rho'^\alpha \, e^{-\rho'}\frac{L_n^{(\alpha)}(\rho) L_n^{(\alpha)}(\rho')}{n+a} \]
Indeed, verifying that it satisfies the differential equation above, we obtain:
\[
\rho\frac{d^2}{d\rho^2}G(\rho,\rho') + (\alpha+1 - \rho)\frac{d}{d\rho}G(\rho,\rho') -a\, G(\rho,\rho') = \sum_{n=0}^\infty \frac{n!}{\Gamma(\alpha+1+n)} \rho'^\alpha \, e^{-\rho'}\, L_n^{(\alpha)}(\rho) L_n^{(\alpha)}(\rho') = \delta(\rho-\rho'),
\]
where we have used the identities
\begin{eqnarray}
\frac{d}{d\rho}L_n^{(\alpha)}(\rho) &=& - L_{n-1}^{(\alpha+1)}(\rho), \\
\rho L_{n-2}^{(\alpha+2)}(\rho) &=& (\alpha+1-\rho) L_{n-1}^{(\alpha+1)}(\rho) - n L_n^{(\alpha)}(\rho)
\end{eqnarray}
and the orthogonality relationship for the Laguerre polynomials:
\[
\sum_{n=0}^\infty \frac{n!}{(\alpha+1)_n}\, L_n^{(\alpha)}(\rho) L_n^{(\alpha)}(\rho') = \Gamma(\alpha+1)\, \rho^{-\alpha}\, e^{\rho}\, \delta(\rho-\rho').
\]
From this, it is clear that eqn.~\eqref{eq:greensolution} is a particular solution to the
differential equation~\eqref{eq:inhomeqn}.
%
\subsection{Integral equation for the spin-singlet case}
Starting from equation \eqref{eq:basicform} we rewrite it into:
\[ \rho k''(\rho) + (1+2\tau-2s - \rho) k'(\rho) - \lambda k(\rho) = Q_\rho k(\rho) \]
Here $\lambda\neq 0,-1,-2,-3,...$ but otherwise arbitrary and serves the purpose of
being able to define a Green function as an expansion in the eigenfunctions. A discussion on the
independence of the results on a particular choice of $\lambda$ is postponed till
section~\ref{sec:determinant}.

The operator $Q_\rho$ is defined as
\begin{equation}
Q_\rho = \frac{y\rho}{1+y\rho} \frac{d}{d\rho} + \tau-z-s-\lambda + \frac{\frac{1}{2} + y(\tau-1-s+\ldots)}{1+y\rho}
\label{eq:opQ}
\end{equation}
With the Green function $G(\rho,\rho')$ defined as in the previous section, the equation for
$k(\rho)$ becomes:
\begin{equation}
k(\rho) = C_1 {}M(\lambda;\alpha+1;\rho) + C_2 U(\lambda;\alpha+1;\rho) + \int_0^\infty d\rho' \, G(\rho,\rho') Q_{\rho'} k(\rho')
\label{eq:homplusint}
\end{equation}
Here, $\alpha=2\tau-2s$ and $M(\lambda;\alpha+1;\rho)$, and $U(\lambda;\alpha+1;\rho)$ are the two
linearly independent solutions to Kummer's equation. The constants $C_1$
and $C_2$ are determined by the boundary conditions at $0$ and $\infty$
\eqref{eq:boundarycond1}-\eqref{eq:boundarycond3}.

The following theorem addresses an important result.
\begin{theorem}
The integral equation
\begin{equation}
k(\rho) = \int_0^\infty d\rho' \, G(\rho,\rho') Q_{\rho'} k(\rho')
\label{eq:integraleqn}
\end{equation}
is a solution to \eqref{eq:basicform} satisfying the boundary conditions
\eqref{eq:boundarycond1}-\eqref{eq:boundarycond3}.
\end{theorem}
The proof of this theorem is contained in Appendix \ref{sec:proofieqn}.
%
\section{\label{sec:determinant}Determinant equation for the eigenvalues}
Starting point is the equation \eqref{eq:integraleqn} which we repeat here for convenience:
\[ k( \rho ) =\int ^{\infty }_{0} \ d\eta \ G( \rho ,\eta ) \ Q_{\eta } \ k( \eta ) \]
and $\lambda$ may take on any complex value except for the eigenvalues of the original
differential equation. To solve equation \eqref{eq:integraleqn}, we assume an expansion
of $k( \rho )$ in associated Laguerre polynomials:
\begin{equation}
k( \rho ) =k_{0} \ L^{( \alpha )}_{0}( \rho ) \ +\ k_{1} \ L^{( \alpha )}_{1}( \rho ) \ +\ \cdots \ =
	\sum ^{\infty }_{n=0} \ k_{n} \ L^{( \alpha )}_{n}( \rho )
\end{equation}
Substituting this expression into the integral equation above, using the well-known recurrence
relations for the associated Laguerre polynomials, and comparing the coefficients of
$L^{( \alpha )}_{n}( \rho )$ and using the lemmas above, after straightforward but tedious
algebra we get
\begin{equation}
\sum _{m=0}^{\infty }\, \frac{L_{m}^{( \alpha )}( \rho )}{m+\lambda}\, \left[\sum _{n=0}^{\infty }\, k_{n}\, \Delta _{m,n}\right] = 0
\label{eq:detequation}
\end{equation}
Here $\Delta$ is the semi-infinite determinant with elements
\begin{equation}
\Delta_{m,n} = ( m+\tau-z-s)\, \delta _{m,n} + \left( yn+\frac{1}{2} + (\tau-1-s+\ldots)y\right)\, \frac{m!}{\Gamma ( \alpha +m+1)}\, I_{m,n} -y\, 
	( n+\alpha ) \frac{m!}{\Gamma ( \alpha +m+1)}\, I_{m,n-1}
\end{equation}
Notice that the determinant does not depend on $\lambda$ as expected. The symbols $I_{m,n}$
are defined in equation \eqref{eq:defimn}. To keep the notation simple we will use $I_{m,n}$
to mean $I^{(\alpha)}_{m,n}$.\\
The determinant can be further simplified by introducing the functions $\overline{I}$ and $\shat$
which we introduce below:
\begin{equation}
\label{eq:defibar}
\overline{I}_{m,n} = \frac{m!}{\Gamma ( \alpha +m+1)}\, I_{m,n}
\end{equation}
Using the $\overline{I}$ the rewritten determinant elements read:
\begin{equation}
\Delta_{m,n} = ( m+\tau-z-s)\, \delta _{m,n} + \left( yn+\frac{1}{2}+(\tau-1-s+\ldots)y\right)\, \overline{I}_{m,n} -y\, ( n+\alpha ) \overline{I}_{m,n-1}
\end{equation}
Next, we use the recurrence relations \eqref{eq:ttrr} for the $\overline{I}$ to finally obtain:
\begin{eqnarray}
\Delta_{m,n} &=& ( m+\tau-z-s) \ \delta _{m,n} -\delta _{m,n-1} -\delta _{m,n-2} -\cdots -\delta _{m,1} -\delta _{m,0} \nonumber \\
&&+ \frac{m!}{\Gamma ( \alpha +m+1)} \ \left( S_{n}^{m} -\left(\frac{1}{2} + (s+1-\tau-\ldots)y\right) \ I_{m,n}\right) \nonumber\\
&\equiv& ( m+1-\zeta) \ \delta _{m,n} -\delta _{m,n-1} -\delta _{m,n-2} -\cdots -\delta _{m,1} -\delta _{m,0} + \shat_{n}^{m}
\label{eq:defdet}
\end{eqnarray}
where the $S^m_n$ and $\shat^m_n$ are defined in equations
\eqref{eq:defsmn} and \eqref{eq:defshat} while $\zeta$ is defined as
\begin{equation}
\zeta = z+s-(\tau-1) \label{eq:defzeta}
\end{equation}
The relevant physics is contained in the variable $\zeta$, which includes the eigenvalues and therefore the
energy levels. The functions $\shat$ include information on the spin-states. Since the Laguerre polynomials
form a linearly independent set, equation \eqref{eq:detequation} has a solution only when
\begin{equation}
\det\, \Delta_{m,n} = 0 \label{eq:deteqndelta}
\end{equation}
Note that the terms in the determinant, as determined by \eqref{eq:defdet}, contain the functions
$I_{m,n}$, either directly or through the definition of the $\shat$ \eqref{eq:defsmn}. When
naively expanding in $1/y\sim\alpha$ logarithms $\ln(1/\alpha)$ of the electromagnetic coupling
constant appear. This follows from the expansion of the $I_{m,n}$ for small $1/y$, see equation
\eqref{eq:imnexpansion}. This differs from the hydrogen atom as studied in \cite{Ish51a,Ish51b}.
Recent calculations on positronium \cite{Adk15} to order $m\alpha^{7}$ show that terms
$\ln(1/\alpha)$ appear.
%
\subsection{\label{sec:dirac}Comparison of the Breit equation to the Dirac equation}
Before analysing \eqref{eq:deteqndelta} to find the eigenvalues, we compare our results
for $J=0$ and $L=0$ (spin singlet case) thus far with the Dirac case ($L=0$). This is
interesting for two reasons. The first is that the exact solution to the
Dirac case is well-known and is a special case of \eqref{eq:basicform}. The eigenfunctions for
the Dirac equation are polynomials in the radial variable ($\rho$). Therefore, we expect that
both the three-term recurrence relation \eqref{eq:heunabc} and the CF
\eqref{eq:ctdfraction} truncate. How this works out is not immediately clear. The second
reason is that the comparison gives valuable insight into the properties of the semi-infinite
determinant that leads to the known solution of the Dirac equation.

In the case of Dirac, the relevant variables read:
\[
s'\equiv 1-\sqrt{1-\alpha ^{2}}, q'=\sqrt{m^{2} -E^{'2}}, y'=\frac{E'+m}{2\alpha q'} =\frac{E'+m}{2\alpha \sqrt{m^{2} -E^{\prime 2}}}
\]
From these redefined variables, it is straightforward to derive
\[
\frac{E'}{m} = \frac{4\alpha ^{2} y^{\prime 2} -1}{4\alpha ^{2} y^{\prime 2} +1}
\]
and
\begin{equation}
\frac{\alpha E'}{q'} = \alpha ^{2} y'-\frac{1}{4y'} = s'( 2-s') y'-\frac{1}{4y'} \equiv z'
\label{eq:defzprime}
\end{equation}
where we have introduced the eigenvalue $z'$. Finally, we express the energy levels in terms
of $z'$:
\[
\frac{E'}{m} =\frac{4\alpha ^{2} y^{\prime 2} -1}{4\alpha ^{2} y^{\prime 2} +1} =\frac{z'}{\sqrt{z^{\prime 2} +\alpha ^{2}}} =\frac{1}{\sqrt{1+\frac{\alpha ^{2}}{z^{\prime 2}}}}
\]
Next, we examine the integral equation \eqref{eq:integraleqn} for $J=0$ and $S=0$
and introduce the operator $\mathfrak{I}$:
\[ k( \rho ) =\int _{0}^{\infty } d\eta\, G( \rho ,\eta )\, Q_{\eta }\, k( \eta ) \equiv \mathfrak{I}\, k( \rho ) \]
\begin{lemma}
The operator $\mathfrak{I}$ acting on the generalized Laguerre polynomials results in:
\[
\mathfrak{I}\, L_{n}^{( \alpha )}( \rho ) = L_{n}^{( \alpha )}( \rho ) - \sum _{m< n}\frac{L_{m}^{( \alpha )}( \rho )}{m+\lambda }\, \left(\shat_{n}^{m} -1\right) -\left( n+1-z-s+\shat_{n}^{n}\right)\frac{L_{n}^{( \alpha )}( \rho )}{n+\lambda } -\sum _{m >n}\frac{L_{m}^{( \alpha )}( \rho )}{m+\lambda }\, \shat_{n}^{m}
\]
\end{lemma}
From this and using properties of the $\shat^m_n$ together with the
expression between $z$ and $y$, we construct the eigenfunctions to the operator
$\mathfrak{I}$:
\begin{equation}
\shat_{n}^{n}\, L_{n-1}^{( \alpha )}( \rho ) - \shat_{n-1}^{n}\, L_{n}^{( \alpha )}( \rho )
\label{eq:linearcomb}
\end{equation}
Before we can study the effect of the operator $\mathfrak{I}$ on the
linear combination \eqref{eq:linearcomb} we need a remarkable property of the
$\shat^m_n$.
%
\begin{lemma}
The $\shat^m_n$ have the properties
\begin{eqnarray}
\shat_{n}^{n} \shat_{n-1}^{m} - \shat_{n-1}^{n} \shat_{n}^{m} &=& 0, \text{when $m>n-1$} \\
\shat_{n}^{n}\left(\shat_{n-1}^{m} -1\right) -\shat_{n-1}^{n} \left( \shat_{n}^{m} - 1 \right) &=& -\frac{m!}{\Gamma ( \alpha +m+1)}\, \frac{I_{m,n}}{\alpha +n}\, \left\{n+1-s-\left[( 2-s) sy-\frac{1}{4y} \right] \right\}, \text{when $m<n$}
\end{eqnarray}
Proof. These relations follow directly from the representations \eqref{eq:shatrepr}.
\end{lemma}
With this property, we are in a position to analyse the effect of the operator $\mathfrak{I}$
on the linear combination \eqref{eq:linearcomb}:
\begin{eqnarray}
&\mathfrak{I} \left[\shat_{n}^{n}\, L_{n-1}^{( \alpha )}( \rho )-\shat_{n-1}^{n}\, L_{n}^{( \alpha )}( \rho )\right] & =
\left[\shat_{n}^{n} \ L_{n-1}^{( \alpha )}( \rho ) -\shat_{n-1}^{n} \ L_{n}^{( \alpha )}( \rho )\right] \nonumber\\
&& -( n+1-z-s) \ \left[\shat_{n}^{n}\frac{L_{n-1}^{( \alpha )}( \rho )}{n-1+\lambda } -\shat_{n-1}^{n}\frac{L_{n}^{( \alpha )}( \rho )}{n+\lambda }\right] \nonumber\\
&& +\left( n+1-s-\left[( 2-s) sy-\frac{1}{4y}\right]\right) \sum _{m< n}\frac{L_{m}^{( \alpha )}( \rho )}{m+\lambda } \ \frac{m!}{\Gamma ( \alpha +m+1)} \ \frac{I_{m,n}}{\alpha +n}
\label{eq:effectOpI}
\end{eqnarray}
Only in the case of Dirac, we recognize the term $( 2-s) sy-\frac{1}{4y} =z$~\eqref{eq:defzprime}.
From this equation, we see directly that for $z=n+1-s$, the function
$\left[\shat_{n}^{n} \ L_{n-1}^{( \alpha )}( \rho ) -\shat_{n-1}^{n}\, L_{n}^{( \alpha )}( \rho )\right]$
is an eigenfunction of the operator $\mathfrak{I}$.

Note that in general this is not true. For example, in the case of the Breit equation
$( 2-s) sy-\frac{1}{4y} \neq z$ and the contributions of the higher order $L_{n}$'s
in \eqref{eq:effectOpI} do not vanish.
%
\subsection{\label{sec:detconv}Convergence and construction of the determinant $\mathbf{\det\, \Delta_{m,n}}$}
The necessary and sufficient conditions that the (semi-)infinite determinant converges
are that (a) the product of the diagonal elements is absolutely convergent and (b) the sum of the
off-diagonal elements is absolutely convergent \cite{Whi20,Ber68}. The sum over
all non-diagonal elements converges absolutely due to the presence of the exponential factor.
The (infinite) product of diagonal elements in \eqref{eq:defdet}, however, does not converge
absolutely.\\

To make the determinant \eqref{eq:defdet} convergent, we are free to multiply
the $m^{\text{th}}$ row in the system of equations \eqref{eq:detequation} for $k_{n}$ by a
constant. This does not change the equation for the $m^{\text{th}}$ row, and this constant can
be chosen differently for each row. We choose the constant such that (a) the product
of diagonal elements is absolutely convergent and (b) the sum of all off-diagonal elements
remains absolutely convergent. A constant that has these properties is
\begin{equation}
\frac{m+1+\left(z+s-\tau+1-\shat_{m}^{m}\right)}{(m+1)^{2}}
\end{equation}
and gives a new determinant $\Delta'_{m,n}$ (with $\zeta=z+s-\tau+1$):
\begin{equation}
\Delta'_{m,n} = \frac{( m+1)^{2} -\left( \zeta-\shat_{m}^{m}\right)^{2}}{( m+1)^{2}}\, \delta _{m,n} +\frac{m+1+\left(\zeta-\shat_{m}^{m}\right)}{(m+1)^{2}}\, ( 1-\delta _{m,n})\left(\shat_{n}^{m} - H\left( n-m-\frac{1}{2}\right)\right)
\label{eq:detprime}
\end{equation}
Here, $H(x)$ is the step function, which is zero for $x<0$ and $1$ otherwise.

Then we get the product of the diagonal elements ($m=n$):
\begin{equation}
\prod _{m=0}^{\infty } \frac{( m+1)^{2} -\left( \zeta-\shat_{m}^{m}\right)^{2}}{( m+1)^{2}} =\prod _{m=0}^{\infty }\left( 1-\frac{\left( \zeta-\shat_{m}^{m}\right)^{2}}{( m+1)^{2}}\right)
\end{equation}
which is absolutely convergent due to \eqref{eq:stildeasymp}. Because of the presence of
the exponential factor in $\shat^m_n$ in \eqref{eq:stildeasymp},
the sum of the off-diagonal elements is absolutely convergent. From this, it follows that the
infinite determinant $\Delta'$ with elements $\Delta'_{m,n}$ is absolutely convergent.

In an explicit form the determinant is given by:
\begin{equation}
\det \Delta'=\begin{vmatrix}
\frac{1-\left( \zeta-\shat_{0}^{0}\right)^{2}}{1^{2}} & \frac{1+\zeta-\shat_{0}^{0}}{1^{2}} \ \left(\shat_{1}^{0} -1\right) & \frac{1+\zeta-\shat_{0}^{0}}{1^{2}} \ \left(\shat_{2}^{0} -1\right) & \frac{1+\zeta-\shat_{0}^{0}}{1^{2}} \ \left(\shat_{3}^{0} -1\right) & \cdots \\[1em]
\frac{2+\zeta-\shat_{1}^{1}}{2^{2}} \ \shat_{0}^{1} & \frac{2^{2} -\left( \zeta-\shat_{1}^{1}\right)^{2}}{2^{2}} & \frac{2+\zeta-\shat_{1}^{1}}{2^{2}} \ \left(\shat_{2}^{1} -1\right) & \frac{2+\zeta-\shat_{1}^{1}}{2^{2}} \ \left(\shat_{3}^{1} -1\right) & \cdots \\[1em]
\frac{3+\zeta-\shat_{2}^{2}}{3^{2}} \ \shat_{0}^{2} & \frac{3+\zeta-\shat_{2}^{2}}{3^{2}} \ \shat_{1}^{2} & \frac{3^{2} -\left( \zeta-\shat_{2}^{2}\right)^{2}}{3^{2}} & \frac{3+\zeta-\shat_{2}^{2}}{3^{2}} \ \left(\shat_{3}^{2} -1\right) & \cdots \\[1em]
\frac{4+\zeta-\shat_{3}^{3}}{4^{2}} \ \shat_{0}^{3} & \frac{4+\zeta-\shat_{3}^{3}}{4^{2}} \ \shat_{1}^{3} & \frac{4+\zeta-\shat_{3}^{3}}{4^{2}} \ \shat_{2}^{3} & \frac{4^{2} -\left( \zeta-\shat_{3}^{3}\right)^{2}}{4^{2}} & \cdots \\[1em]
\vdots  & \vdots  & \vdots  & \vdots  & \ddots 
\end{vmatrix}
\label{eq:detexplicit}
\end{equation}
The determinant $\det\Delta'$ is the basis for our procedure for determining the
first and second order coefficients of the expansion for the eigenvalues $z_n$.
%
\section{\label{sec:smally}Solving the determinant by numerical methods}
To verify that the eigenvalues based on the determinant~\eqref{eq:detexplicit} are consistent
with the eigenvalues determined by the CF~\eqref{eq:eigenvalues}, we numerically
check this. Direct numerical analysis of~\eqref{eq:detexplicit} is challenging due to the asymptotic
behavior of the $I_{nm}$ for large $m,n$ (see \lemref{lem:asymptotic}). In the physical
region, the variable $y\sim 38000$ and the elements of the determinant will be small only for very
large $m,n$. The consequence is that to get accurate results, large dimensions of the determinant
are needed to get a decent approximation. Therefore, we base the calculations on treating the
variables $z$ and $y$ as independent, which enables us to compare results for small $y\in \{1, 10, 100\}$.
To cope with challenges in the numerical calculation, we exploit three different methods: (a) based
on the direct calculation, (b) based on an equivalent forward recursion relation, and (c) based on
the CF~\eqref{eq:eigenvalues}. This enables us to verify the equivalence and consistency
of the Green function method and the CF method.
\subsection{The direct method}
When calculating the determinant in a direct way from~\eqref{eq:detexplicit}, we can simplify the
calculation tremendously by exploiting relation~\eqref{eq:shatsymmetry} while using the condensation
method by Dodgson \cite{Tur60}. Due to~\eqref{eq:shatsymmetry} we are left with a tri-band matrix after
the first reduction, and each subsequent reduction (condensation) leaves us with a tri-band
structure.
\subsection{Forward recursion method}
In the forward recursion method, we reduce a determinant of size $N$ to a tri-band structure.
Again, using the relations~\eqref{eq:shatsymmetry} and~\eqref{eq:shatasymmetry} the following
equivalent determinant is derived:
\begin{equation}\label{eq:detfinite}
\Delta_{n,N} = \left| \begin{array}{ccccccc}
a_n     & c_n & 0 & 0 & \cdots & 0& 0 \\
b_n     & a_{n+1} & c_{n+1} & 0 & \cdots & 0& 0 \\
0         & b_{n+1} & a_{n+2} & c_{n+2} & \cdots & 0& 0 \\
\vdots & \vdots & \vdots & \vdots & \ddots & \vdots & \vdots \\
0 & 0    & 0 & 0 & \cdots & a_{N-1} & c_{N-1} \\
0 & 0    & 0 & 0 & \cdots & b_{N-1} & \frac{N^2 - \left(\zeta-\shat^{N-1}_{N-1}\right)^2}{N^2}
\end{array} \right|, \,\, \det\Delta' = \Delta_{1,N}
\end{equation}
The coefficients $a_n$, $b_n$, and $c_n$ are given by:
\begin{subequations}
\label{eq:detrecabc}
\begin{eqnarray}
&&a_n = \frac{n+\zeta-\shat_{n-1}^{n-1}}{n^{2}} \ \left(( n+1-\zeta) \ \left( 1+\frac{\shat_{n-1}^{n}}{\shat_{n}^{n}} \ \frac{\shat_{n}^{n-1} -1}{\shat_{n}^{n} -1}\right) +\shat_{n-1}^{n-1} -1-\left(\shat_{n}^{n-1} -1\right) \ \frac{\shat_{n-1}^{n}}{\shat_{n}^{n}}\right) \\
&&b_n = -\frac{n+1+\zeta-\shat_{n}^{n}}{(n+1)^{2}} \ \frac{\shat_{n-1}^{n}}{\shat_{n}^{n}} \ ( n+1-\zeta) \\
&&c_n = -\frac{n+\zeta-\shat_{n-1}^{n-1}}{n^{2}} \ \frac{\shat_{n}^{n-1} -1}{\shat_{n}^{n} -1} \ ( n+2-\zeta)
\end{eqnarray}
\end{subequations}
The determinant~\eqref{eq:detfinite} is recursively calculated with:
\begin{equation}
\Delta_{n,N} = a_n\ \Delta_{n+1,N} - b_n\ c_n\ \Delta_{n+2,N}
\label{eq:detrecursion}
\end{equation}
\subsection{Results}
For relatively small values of $y=1,10,100$, we compare the 'direct' (a) and the 'continuous fraction' (c)
methods. In this region, we expect that the $\shat$-functions approach zero fast due to their asymptotic
behavior as formulated in \lemref{lem:asymptotic}. The results are shown in
Tables~\ref{tab:numy1}-\ref{tab:numy100}. In establishing the results, the calculations are based on
the Python package {\tt mpmath} \cite{mpmath}, and 50 digits of precision are used when using both methods.
The direct calculation for a 200x200 determinant already yields over 20 digits of accuracy for $y=1$ compared
to the CF method, while an 800x800 determinant yields over 8 digits of accuracy for $y=100$
illustrating that achieving high accuracy for $y>100$ is cumbersome based on the 'direct' method.\\

To analyze the numerical stability and convergence of the algorithm locating the zeros for varying scales of $y$,
in Tables~\ref{tab:numy1}-\ref{tab:numy100} the variable $y$ is varied independently from $z$ and $\zeta$.
This is achieved through the equations \eqref{eq:heunabc}, and through the $\shat$-functions in \eqref{eq:detrecabc}.
\begin{table}[H]
\begin{tabular}{llllll}
Size (determinant) & Digits & Method & $z$ (determinant) & Digits & $z$ (CF) \\ \hline
50 & 50 & Direct & $\texttt{\underline{1.1727693519545} 30126415377}$ & 50 & $\texttt{\underline{1.1727693519548} 42311051307}$ \\
100 & 50 & Direct &$\texttt{\underline{1.172769351954842309} 142413}$ & 50 & $\texttt{\underline{1.172769351954842311} 051307}$ \\
200 & 50 & Direct & $\texttt{\underline{1.17276935195484231105125} 0}$ & 50 & $\texttt{\underline{1.17276935195484231105130} 7}$
\end{tabular}
\caption{\label{tab:numy1}Numerical comparison for $N=1, y=1$ based on the (a) direct method and (c) CF method.}
\end{table}
\begin{table}[H]
\begin{tabular}{llllll}
Size (determinant) & Digits & Method & $z$ (determinant) & Digits & $z$ (CF) \\ \hline
100 & 50 & Direct & $\texttt{\underline{1.03618496} 1647506905861119}$ & 50 & $\texttt{\underline{1.03618498} 0978793013999313}$ \\
200 & 50 & Direct & $\texttt{\underline{1.03618498090} 8419212520244}$ & 50 & $\texttt{\underline{1.03618498097} 8793013999313}$ \\
400 & 50 & Direct & $\texttt{\underline{1.03618498097876} 3675635578}$ & 50 & $\texttt{\underline{1.03618498097879} 3013999313}$ \\
800 & 50 & Direct & $\texttt{\underline{1.0361849809787930137} 12053}$ & 50 & $\texttt{\underline{1.0361849809787930139} 99313}$
\end{tabular}
\caption{\label{tab:numy10}Numerical comparison for spin-singlet $N=1, y=10$ based on the (a) direct method and (c) CF method.}
\end{table}
\begin{table}[H]
\begin{tabular}{llllll}
Size (determinant) & Digits & Method & $z$ (determinant) & Digits & $z$ (CF) \\ \hline
100 & 50 & Direct & $\texttt{\underline{1.004651} 179004409538463433}$ & 250 & $\texttt{\underline{1.004655} 369597654872180760}$ \\
200 & 50 & Direct & $\texttt{\underline{1.0046548} 38709008473502799}$ & 250 & $\texttt{\underline{1.0046553} 69597654872180760}$ \\
400 & 50 & Direct & $\texttt{\underline{1.00465533} 5599754027941414}$ & 250 & $\texttt{\underline{1.00465536} 9597654872180760}$ \\
800 & 50 & Direct & $\texttt{\underline{1.0046553687} 69331040681791}$ & 250 & $\texttt{\underline{1.0046553695} 97654872180760}$
\end{tabular}
\caption{\label{tab:numy100}Numerical comparison for spin-singlet $N=1, y=100$ based on the (a) direct method and (c) CF method.}
\end{table}
For a larger value of $y$, we compare results between all three methods for physical values of $y\sim37537$.
Using the CF method, the calculations need to be carried out with 600 digits to cope with
large cancellations. We employ the 'direct' method up to sizes 8192x8192, achieving 9-digit accuracy. To
obtain larger accuracies, we utilize the 'forward recursion' method based on~\eqref{eq:detfinite}. The
results are shown in Table~\ref{tab:numyz}. The latter method provides over 15 digits of accuracy.
\begin{table}[H]
\begin{tabular}{llllll}
Size (determinant) & Digits & Method & $z$ (determinant) & Digits & $z$ (CF) \\ \hline
100 & 50 & Direct & $\texttt{\underline{0.99999993} 27931242734964910}$ & 600 & $\texttt{\underline{0.99999999} 68218806697002054}$ \\
200 & 50 & Direct & $\texttt{\underline{0.99999996} 53589113318220221}$ & 600 & $\texttt{\underline{0.99999999} 68218806697002054}$ \\
400 & 50 & Direct & $\texttt{\underline{0.99999998} 16445916357823509}$ & 600 & $\texttt{\underline{0.99999999} 68218806697002054}$ \\
800 & 50 & Direct & $\texttt{\underline{0.99999998} 97009655033995428}$ & 600 & $\texttt{\underline{0.99999999} 68218806697002054}$ \\
1200 & 50 & Direct & $\texttt{\underline{0.999999992} 3336934079238152}$ & 600 & $\texttt{\underline{0.999999996} 8218806697002054}$ \\
2000 & 50 & Direct & $\texttt{\underline{0.999999994} 3828846189827440}$ & 600 & $\texttt{\underline{0.999999996} 8218806697002054}$ \\
4096 & 50 & Direct & $\texttt{\underline{0.9999999958} 62119578967378}$ & 600 & $\texttt{\underline{0.9999999968} 218806697002054}$ \\
8192 & 50 & Direct & $\texttt{\underline{0.9999999966} 23488391242820}$ & 600 & $\texttt{\underline{0.9999999968} 218806697002054}$ \\
16384 & 25 & Recursion & $\texttt{\underline{0.9999999967} 21711325628515}$ & 600 & $\texttt{\underline{0.9999999968} 218806697002054}$ \\
65536 & 25 & Recursion & $\texttt{\underline{0.999999996818} 832675025857}$ & 600 & $\texttt{\underline{0.999999996821} 8806697002054}$ \\
131072 & 25 & Recursion & $\texttt{\underline{0.99999999682165} 4233775826}$ & 600 & $\texttt{\underline{0.99999999682188} 06697002054}$ \\
262144 & 25 & Recursion & $\texttt{\underline{0.999999996821873} 839215586}$ & 600 & $\texttt{\underline{0.999999996821880} 6697002054}$ \\
524288 & 25 & Recursion & $\texttt{\underline{0.9999999968218806} 12157009}$ & 600 & $\texttt{\underline{0.9999999968218806} 697002054}$ \\
1048576 & 25 & Recursion & $\texttt{\underline{0.999999996821880669} 615008}$ & 600 & $\texttt{\underline{0.999999996821880669} 7002054}$
\end{tabular}
\caption{\label{tab:numyz}Numerical comparison for spin singlet $N=1, \frac{1}{2}\alpha^2 y=z$ of three
methods for determining the eigenvalues: (a) direct method, (b) forward recursion, and (c) CF.}
\end{table}
Although it does not provide a rigorous proof, the numerical results for obtaining the eigenvalue $z$
from the CF~\eqref{eq:eigenvalues} or from the determinant~\eqref{eq:detexplicit} are
consistent.

%
\section{Conclusion\label{sec:conclusion}}
In this work, we have presented two different approaches to determine the eigenvalues and
energy levels for the Breit equation for the spin singlet ($S=0$) and spin triplet ($S=1$
and $J=0$ or $J=L$). The continued fraction (CF) approach enables the calculation
of the energy levels to a large accuracy. The outcome corresponds to the results known in
the literature \cite{Sco92,Yam12,Tur24}. This includes the reproduction of the spurious
state found in the neighborhood of $\rho\sim 1/y$ \cite{Pat19}. The Green function
approach yields results that are consistent with the CF approach, and offers an alternative
analytic condition determining the energy levels.

While the CF approach lends itself well to solving the problem numerically, it
is very difficult to obtain an analytic expansion of the energy levels in terms of the
electromagnetic fine-structure constant $\alpha^2$. As an alternative to this approach, we presented
a novel method to obtain such an expansion. This novel method is based on reformulating the
eigenvalue problem for a differential equation into one that involves an integral equation.
Expanding the eigenfunctions in associated Laguerre polynomials, we arrived at a condition for
a semi-infinite determinant. This condition is similar to the case of the
CF and determines the eigenvalues. This seems to suggest an expansion in $\alpha$ and
$\ln(1/\alpha)$ which is consistent with recent theoretical calculations \cite{Adk15,Adk22}.
Although we haven't succeeded in exploiting this condition to obtain an explicit expansion
in $\alpha^{2}$, this may be possible in future work. Furthermore, we have shown that the
integral equation and eigenvalues reduce to Dirac's case for a suitable mapping of the
parameters.\\

The functions $\shat$ play an essential role in the analysis of the integral equation.
We found many properties of these functions that ease the manipulation of the semi-infinite
determinant. Not only do they include information on the spin-states (through $\omega$)
and their properties are crucial in extracting the correct eigenvalues in the Dirac case.
In addition, their properties enable us to derive a forward recursion for the semi-infinite
determinant \eqref{eq:detrecursion} from which the eigenvalues can be calculated to high
precision.\\

The results obtained in this work using the CF match the known results in the literature
($N=1,2,3$ in \cite{Sco92}) to high accuracy. We have extended this calculation to $N=4,5$.
Both approaches presented in this work yield an analytical expression for the eigenvalues
as the zeros of a continued fraction or the zeros of a semi-infinite determinant. Moreover,
analytical expressions for the corresponding eigenfunctions have been provided in terms
of an expansion in associated Laguerre polynomials and coefficients that are given by
a three-term recurrence relation.\\
Using either the CF approach or the Green function approach it is not difficult to
include the complete photon exchange interaction (Breit operator) and obtain the proper
ordering of the $2{}^1P_{1}$ and $2{}^1S_{0}$ states.

\appendix
%
\section{\label{sec:imnproperties}The special functions $\bm I_{\bm{m,n}}$ and $\overline{\bm I}_{\bm{m,n}}$}
The functions $I_{m,n}$ are central to our approach for calculating the second-order
contributions to the determinant \eqref{eq:detexplicit}. Their origin is the $1+y\rho$
factor in the $Q$-operator \eqref{eq:opQ} which occurs inside the kernel of the
integral equation \eqref{eq:integraleqn}. In this section, we list important
properties of the $I_{m,n}$.

\subsection{Definition and simple properties}
\begin{definition}
The function $I_{m,n}$ is defined as:
\begin{equation}
\label{eq:defimn}
I^{(\alpha,\beta)}_{m,n}( y) =\int ^{\infty }_{0}\frac{d\rho }{1+y\rho } \rho ^{\alpha }\, e^{-\rho }\, L^{( \alpha )}_{m}( \rho )\, L^{( \beta )}_{n}( \rho )
\end{equation}
To keep the notation simple we will use $I_{m,n}$ to mean $I^{(\alpha)}_{m,n} = I^{(\alpha,\alpha)}_{m,n}$.
\end{definition}
From the orthogonality of the Laguerre polynomials, the rational function $(1+y\rho)^{-1}$
has the expansion where the $I_{m,n}$ occur as its coefficients:
\begin{corollary}
The expansion of the reciprocal of $1+y\rho$ in Laguerre polynomials is
\begin{equation}
\frac{1}{1+y\rho}\, L_{m}^{(\alpha)}(\rho) = \sum_{k=0}^\infty \frac{k!}{\Gamma(\alpha+k+1)}\, I_{m,k}(y)\, L_{k}^{(\alpha)}(\rho)
\label{eq:ratioexpansion}
\end{equation}
and the special case:
\begin{equation}
\frac{1}{1+y\rho} = \sum_{k=0}^\infty \frac{k!}{\Gamma(\alpha+k+1)}\, I_{0,k}(y)\, L_{k}^{(\alpha)}(\rho)
\label{eq:ratioexpansionsimple}
\end{equation}
\end{corollary}
From its definition \eqref{eq:defimn} it is clear that the $I_{m,n}$ possesses the property:
\begin{corollary}
The $I_{m,n}^{(\alpha)}$ is symmetric in its indices $m,n$:
\begin{equation}
I_{m,n} = I_{n,m}
\label{eq:imnsymm}
\end{equation}
\end{corollary}
An important property of \eqref{eq:defimn} is that it factorizes as expressed by the following corollary:
\begin{corollary}
When $n\leq m$ the $I_{m,n}^{(\alpha,\beta)}$ can be written as a product of two factors:
\begin{equation}
I_{m,n}^{(\alpha,\beta)} = L_n^{(\beta)}(-1/y)\, I_{m,0}
\label{eq:symimn}
\end{equation}
and a similar identity holds when $m\leq n$ and $\alpha=\beta$. The result follows from integrating
\eqref{eq:defimn} by parts, the orthogonality of the Laguerre polynomials, and using the identity
\[ \frac{1}{1+y\rho}\rho^m = \frac{1}{y} \left( 1 - \frac{1}{1+y\rho} \right)\rho^{m-1}. \]
\end{corollary}
For convenience, the function $\overline{I}_{m,n}$ is introduced in the following definition.
\begin{definition}
The function $\overline{I}_{m,n}$ is defined as:
\begin{equation}
\label{eq:defibarmn}
\overline{I}^{(\alpha,\beta)}_{m,n}( y) = \frac{m!}{\Gamma(m+\alpha+1)}\; I^{(\alpha,\beta)}_{m,n}( y)
\end{equation}
\end{definition}
\begin{corollary}
From its definition \eqref{eq:defimn} and the property \eqref{eq:symimn} we have for $m<n$
\begin{equation}
\ibar^{(\alpha,\alpha+1)}_{m,n} = 1 - y(\alpha+1)\; \frac{m!}{(\alpha+1)_m}\; L_m^{(\alpha)}(-1/y)\; \ibar^{(\alpha+1)}_{0,n}(y)
\label{eq:iaa1}
\end{equation}
\end{corollary}
%
\subsection{Orthogonality relations}
The definition \eqref{eq:defimn} suggests investigating the orthogonal polynomials corresponding to
the weight function $w(\rho)$ and innerproduct:
\begin{equation}
\label{eq:weight}
w(\rho) = \frac{1}{1+y\rho}\; \rho^\alpha\; e^{-\rho}, \qquad <f,g> \equiv \int_{0}^{\infty} d\rho\; w(\rho)\; f(\rho)\; g(\rho)
\end{equation}
and is captured in the following theorem.
\begin{theorem}
A set of independent polynomials $f_n(\rho)$ that are orthogonal with respect to the weight function \eqref{eq:weight} are:
\begin{equation}
(a)\qquad f_{0}(\rho) = L_{0}^{(\alpha)}(\rho), \qquad f_{n}(\rho) = L_{n}^{(\alpha)}(\rho) - \frac{I_{n,0}^{(\alpha)}}{I_{n-1,0}^{(\alpha)}}\; L_{n-1}^{(\alpha)}(\rho)
\end{equation}
The polynomials $f_{n}(\rho)$ are normalised such that the coefficient of the highest order Laguerre functions is 1.\\
Moreover, the polynomials have the properties:
\[
(b)\qquad <f_{m},f_{n}> = \delta_{m,n}
\begin{cases}
I_{0,0}^{(\alpha)} & \text{when $n=0$} \\[1em]
\frac{I_{n-1,0}^{(\alpha)}}{I_{n,0}^{(\alpha)}}\; \frac{1}{y}\frac{\Gamma(\alpha+n)}{n!} & \text{when $n>0$}
\end{cases}
\]
\[
(c) \qquad f_{0}(-1/y) = I_{0,0}^{(\alpha)} \quad\text{when $n=0$, and}\quad f_{n}(-1/y)=\frac{1}{I_{n-1,0}^{(\alpha)}}\; \frac{1}{y}\; \frac{\Gamma(\alpha+n)}{n!}\quad \text{when $n>0$}
\]
\end{theorem}
%
\subsection{Explicit representation}
Before we can derive an explicit representation of the $I_{m,n}$, we need to utilize the
representation of $(1+y\rho)^{-1}$ as a contour integral.
\begin{lemma}
The rational function $(1+y\rho)^{-1}$ has the following Mellin-Barnes contour integral
representation that is valid for all $y$ and $\rho$:
\begin{equation}
\frac{1}{1+y\rho} = -\frac{1}{2\pi i}\int_{\gamma-i\infty}^{\gamma+i\infty}d\sigma\;(y\rho)^{1+\sigma}\Gamma( 1+\sigma)\Gamma(-\sigma)
\label{eq:mbratio}
\end{equation}
where the points $\sigma=-1,0,1,\dots$ lie to the right of the contour, while the points
$\sigma=-2,-3,\dots$ lie to the left of the contour: $-2<\Re(\gamma)<-1$.
\begin{proof}
The result follows immediately from applying the residue theorem to \eqref{eq:mbratio} and
closing the contour either to the left or to the right, depending on whether $|y\rho| >> 1$
or $|y\rho| << 1$ respectively.
\end{proof}
\end{lemma}
\begin{theorem}
The $I_{0,n}$ can be expressed in Kummer's $U$-function:
\begin{equation}
I_{0,n}(y) = \int_0^\infty \frac{d\rho}{1+y\rho}\, \rho^\alpha\, e^{-\rho}\, L_n^{(\alpha)}(\rho) = \Gamma(n+\alpha+1)\frac{1}{y}\, U(n+1; 1-\alpha; 1/y)
\label{eq:ion}
\end{equation}
\begin{proof}
The result follows immediately from substituting \eqref{eq:mbratio} into \eqref{eq:defimn} and
an application of the residue theorem at the points $\sigma=-2,-3,\dots$. Using the well-known relation
that expresses the Kummer $U$-function in the confluent hypergeometric function,
see e.g. formula (13.1.3) in \cite{AS70}, we obtain \eqref{eq:ion}.
\end{proof}
\end{theorem}
\begin{corollary}
For indices equal to zero, we have the identity:
\begin{equation}
I_{0,0} = \Gamma(\alpha+1)\, y^{-\alpha-1}\, e^{1/y}\, \Gamma(-\alpha,1/y)
\end{equation}
where $\Gamma(-\alpha,1/y)$ is the incomplete gamma-function.
\end{corollary}
A useful representation for the $I_{m,n}$ is as a Mellin-Barnes contour integral.
\begin{theorem}
The Mellin-Barnes representation of $I_{0,n}^{(\alpha)}$ is
\begin{equation}
\frac{1}{2\pi i} \frac{1}{n!}\int_{\gamma}d\sigma\;y^{1+\sigma}\Gamma( 2+\sigma) \Gamma(\alpha+2+\sigma)\Gamma(n-1-\sigma)
\label{eq:mbrepr}
\end{equation}
and the contour $\gamma$ separates the poles of $\Gamma(\sigma+2)\Gamma(\alpha+2+\sigma)$,
from the poles of $\Gamma(n-1-\sigma)$.
\begin{proof}
From the definition of the $I_{m,n}$ \eqref{eq:defimn} and the Mellin-Barnes representation of
$(1+y\rho)^{-1}$ (see \eqref{eq:mbratio}) the result immediately follows.
\end{proof}
\label{thm:mellinbarnes}
\end{theorem}
%
\subsection{Expansions for large argument and indices}
Here we study the behavior of the $I_{m,n}(y)$ for both large $m$ and $n$ in the limit to
infinity, and as a series expansion for small $1/y$. Both cases are treated in the next
two theorems
\begin{theorem}
An expansion of $I_{0,n}$ in $1/y$ and fixed $n$ is given by:
\begin{equation}
I^{(\alpha)}_{0,n}(y) = \frac{\Gamma(\alpha)}{y}\, \left( {}_1F_1(n+1; 1-\alpha; 1/y) - y^{-\alpha}\; \Gamma(1-\alpha)\frac{(\alpha+1)_n}{n!}\; {}_1F_1(n+\alpha+1;\alpha+1;1/y) \right)
\label{eq:imnexpansion}
\end{equation}
\begin{proof}
Starting from equation \eqref{eq:ion} and expressing Kummer's $U$-function as a linear combination of
confluent hypergeometric functions ${}_1F_1(\dots)$ the result follows.
\end{proof}
\end{theorem}
\begin{corollary}
In the case of the spin-singlet the expansion of $I_{0,n}$ in $1/y$ is given by:
\begin{eqnarray}
I^{(\alpha)}_{0,n}(y) &=& \frac{\Gamma(\alpha-1)}{y}\, \left\{\alpha-1 - \frac{n+1}{y} - \frac{(n+1)(n+2)}{[2(2-\alpha)y]}\, \frac{1}{y}\, {}_{2} F_{2}( 1,n+3;3,3-\alpha ;1/y) \right. \nonumber\\*
&& \left. + \frac{2\ \Gamma(3-\alpha)}{[2(2-\alpha)y]}\, \left[\frac{( \alpha +1)_{n}}{n!}\right]\, \left(\frac{1}{y}\right)^{\alpha-1}\, {}_{1} F_{1}( n+\alpha +1;1+\alpha ;1/y)\right\}
\label{eq:imnasymp}
\end{eqnarray}
For the spin-singlet $\tau=1$ leading to $\alpha=2-2s$ \eqref{eq:sigmaparameters}. To prevent
the term $2-\alpha$ from occurring in the denominator of the hypergeometric function, the
following identity is utilized:
\[ {}_1F_1(n+1;1-\alpha;1/y) = 1 - \frac{n+1}{(\alpha-1)y} - \frac{(n+1)(n+2)}{2(\alpha-1)y}\frac{2-s}{z}\; {}_2F_2(1,n+3;3,3-\alpha;1/y) \]
where we have used the definitions of $z$ \eqref{eq:zdefinition} and $s$ \eqref{eq:staudefinition} to rewrite
\[ (2-\alpha)y = 2s y = \frac{(2\tau-s)2s y}{2\tau-s} = \frac{z}{2\tau-s} \]
\end{corollary}
The behavior of the $I_{m,n}$ for large $m$ and $n$ is addressed in the following theorem.
\begin{theorem}\label{lem:asymptotic}
For fixed $m$ and in the limit $n\to\infty$ the $I_{m,n}$ behaves as:
\[
(i)\,\, I_{m,n} \sim \sqrt{\pi}\, y^{-\frac{1}{2}\alpha-\frac{3}{4}}\, e^{\frac{1}{2y}}\, L_m^{(\alpha)}(-1/y)\, n^{\frac{1}{2}\alpha-\frac{1}{4}}\, e^{-2\sqrt{n/y}}
\]
and for fixed $n$ and in the limit $m\to\infty$ the $I_{m,n}$ behaves as:
\[
(ii)\,\, I_{m,n} \sim \sqrt{\pi}\, y^{-\frac{1}{2}\alpha-\frac{3}{4}}\, e^{\frac{1}{2y}}\, L_n^{(\alpha)}(-1/y)\, m^{\frac{1}{2}\alpha-\frac{1}{4}}\, e^{-2\sqrt{m/y}}
\]
while for $m=n$ and $m\to\infty$ the function $I_{m,m}$ has the behavior:
\[
(iii)\,\, I_{m,m} \sim \frac{1}{2\sqrt{y}}\, m^{\alpha-\frac{1}{2}}
\]
\begin{proof}
The result for the cases $(i)$ and $(ii)$ is implied by equation \eqref{eq:ion} and using the
asymptotic expansion for Kummer's $U$-function and the modified Bessel function $K_\nu$,
see e.g. equations (3.8) in \cite{Tem13} and 8.451-6 in \cite{Gra65}:
\[ U(n+1;1-\alpha;1/y) \sim \frac{\sqrt{\pi}}{n!}\; y^{-\frac{1}{2}\alpha+\frac{1}{4}}\; n^{-\frac{1}{2}\alpha-\frac{1}{4}}\; e^{1/2y}\; e^{-2\sqrt{n/y}} \]
Using that for large $n$ the ratio of two Gamma-functions $\Gamma(n+\alpha+1)/n!\sim n^\alpha$ we obtain
\[ I_{0,n} \sim \sqrt{\pi}\; y^{-\frac{1}{2}\alpha-\frac{3}{4}}\; n^{\frac{1}{2}\alpha-\frac{1}{4}}\; e^{1/2y}\; e^{-2\sqrt{n/y}} \]
The third case $(iii)$ is established by using equation (24) in \cite{Dea13} or equation (43)
in \cite{Bor07} for the leading term of the expansion of Laguerre polynomials for large $n$ and reads
\begin{equation}
L_n^{(\alpha)}(-1/y) = \frac{1}{2\sqrt{\pi}}\frac{\Gamma(\alpha+n+1)}{n!}\, e^{-1/2y}\, (\kappa/y)^{-\alpha/2-1/4}\, e^{2\sqrt{n/y}}\, \left[ 1 + {\cal O}(n^{-1/2}) \right]
\end{equation}
where $\kappa=\frac{1}{2}(\alpha+1)+n$. Substituting this expression in the result for either
$(i)$ or $(ii)$ gives the stated result.
\end{proof}
\end{theorem}
%
\subsection{Generating function}
An interesting curiosity is that the $I_{0,0}^{(\alpha)}$ is its own generating function which is the
result of the next theorem. A formal definition of generating functions can be found in \cite{Gou83}.
\begin{theorem}
The function $G(w)$ defined for $|w|<1$ by
\begin{equation}
G(w) = (1-w)^{\alpha -\beta } \ I_{0,0}^{(\alpha)}(y( 1-w))
\end{equation}
is the generating function of the $I_{0,n}^{(\alpha,\beta)}$:
\begin{equation}
G(w) = \sum _{k=0}^{\infty } I_{0,k}^{(\alpha, \beta )}(y)\ w^{k}
\label{eq:genimn}
\end{equation}
\begin{proof}
Starting from \eqref{eq:genimn}, using \eqref{eq:defimn}, replacing the sum by the generating function
of the Laguerre polynomials, see 8.975-1 in \cite{Gra65}
\[ \sum_{n=0}^\infty L_n^{(\alpha)}(\rho)\, w^n = (1-w)^{-\alpha-1}\, \exp\left(-\frac{\rho w}{1-w}\right) \]
and performing the integration we have:
\begin{eqnarray*}
\sum _{k=0}^{\infty} I_{0,k}^{( \alpha ,\beta )}(y)  w^{k} &=& \int _{0}^{\infty }\frac{d\rho }{1+y\rho }  \rho ^{\alpha }  e^{-\rho }  \sum _{k=0}^{\infty } L_{k}^{( \beta )}( \rho )  w^{k} =( 1-w)^{-\beta -1}  \int _{0}^{\infty }\frac{d\rho }{1+y\rho }  \rho ^{\alpha }  e^{-\rho }  \exp\left(\frac{-w\rho }{1-w}\right) \\*
&=&( 1-w)^{-\beta -1}  \int _{0}^{\infty }\frac{d\rho }{1+y\rho }  \rho ^{\alpha }  \exp\left(\frac{-\rho }{1-w}\right)
= ( 1-w)^{\alpha -\beta }\, I_{0,0}( y( 1-w))
\end{eqnarray*}
\end{proof}
\end{theorem}
%
\subsection{\label{sec:ttrr}Recurrence relations}
\begin{theorem}
The $I_{m,n}$ satisfy the following three term recurrence relation:
\begin{equation}
\label{eq:ttrr}
( n+1)\, I_{m,n+1} - ( 2n+\alpha +1+1/y)\, I_{m,n} + ( n+\alpha )\, I_{m,n-1} = -\frac{1}{y}\frac{\Gamma ( \alpha +n+1)}{n!} \delta _{m,n}
\end{equation}
\begin{proof}
Multiplying \eqref{eq:defimn} by $1/y$ and partial fractioning we get
\[
\frac{1}{y}\, I_{m,n} = \int_0^\infty\, \left(\frac{1}{y\rho} - \frac{1}{1+y\rho}\right)\, \rho^\alpha e^{-\rho}\, \left(\rho\, L_m^{(\alpha)}(\rho)\right)L_n^{(\alpha)}(\rho)
= \frac{1}{y}\frac{\Gamma(m+\alpha+1)}{m!}\, \delta_{m,n} - \int_0^\infty\, \frac{1}{1+y\rho}\, \rho^\alpha e^{-\rho}\, L_m^{(\alpha)}(\rho)\left(\rho\, L_n^{(\alpha)}(\rho)\right)
\]
The result then follows using the recurrence relations for the Laguerre polynomials,
see equation 8.971-6 in \cite{Gra65}, and identifying the $I_{m,n}$'s from their definition \eqref{eq:defimn}.
\end{proof}
\end{theorem}
%
\subsection{\label{sec:imnsums}Various summation formulae}
In this section, we list various sums involving the $I_{m,n}$ that occur often in our calculations:
\begin{theorem}
\begin{eqnarray}
&&(i) \sum_{m=0}^{\infty }\frac{m!}{(\alpha+1)_{m}}\, I_{0,m} = \Gamma(\alpha+1) - \alpha y\, I_{0,0} \label{eq:isumsingle} \\
&&(ii) \sum_{m=0}^{\infty} \frac{m!}{(\alpha+1)_{m}}\, I_{0,m}\, I_{0,m} = \frac{1}{y}\Gamma(\alpha+1) \left(\Gamma(\alpha+1) - \alpha y\, I_{0,0} - I_{0,0} \right)\label{eq:isumdouble}
\end{eqnarray}
\begin{proof}
Case $(i)$. The first part is derived by exploiting the well-known sum \cite{Han75}
\[ \sum_{n=0}^\infty\, \frac{n!}{(\alpha+1)_n}\, L_n^{(\alpha)}(\rho) = \frac{\alpha}{\rho} \]
and switching the sum and integral, we have
\[
\sum_{m=0}^{\infty }\frac{m!}{(\alpha+1)_{m}}\, I_{0,m} = \int_0^\infty\, \frac{d\rho}{1+y\rho}\, \rho^\alpha\, e^{-\rho}\, \frac{\alpha}{\rho}
= \alpha y\, \int_0^\infty\, \frac{d\rho}{1+y\rho}\, \rho^\alpha\, e^{-\rho}\, \frac{1}{y\rho}
= \alpha\, ( \Gamma(\alpha) - y\, I_{0,0} )
\]
which establishes the result.\\
Case $(ii)$. The completeness relation for Laguerre polynomials reads:
\[ \sum_{m=0}^\infty \frac{m!}{(\alpha+1)_m}\, L_m^{(\alpha)}(\rho)\, L_m^{(\alpha)}(\rho') = \Gamma(\alpha+1) \rho^{-\alpha}\, e^{-\rho}\, \delta(\rho-\rho') \]
where $\delta(\rho-\rho')$ is the Dirac delta-function. Substituting the defining equation \eqref{eq:defimn}
for $I_{0,m}$ into the sum and integrating by parts gives:
\[
\sum_{m=0}^{\infty} \frac{m!}{(\alpha+1)_{m}}\, I_{0,m}\, I_{0,m} = \Gamma(\alpha+1)\, \int_0^\infty\, \frac{d\rho}{(1+y\rho)^2}\, \rho^\alpha\, e^{-\rho}
= \Gamma(\alpha+1)\, \int_0^\infty\, \frac{d\rho}{1+y\rho} \left( \frac{\alpha}{y\rho} - \frac{1}{y}\right)\, \rho^\alpha\, e^{-\rho}
\]
After partial fractioning the first term, the result follows.
\end{proof}
\end{theorem}

%
\section{\label{sec:proofieqn}Proof of the integral equation for $k(\rho$)}
To proof that both constants $C_1$ and $C_2$ in \eqref{eq:homplusint} are zero we need
knowledge on the behaviour of
\[ \int_0^\infty d\rho'\; G(\rho,\rho') Q_{\rho'}{\rho'}^\sigma \]
for $\rho\sim 0$ and $\rho\to\infty$. Before we can address this question
two lemma's involving Lagurre polynomials need to be established.
\begin{lemma}
\begin{equation}
\int_0^\infty \frac{d\rho}{1+y\rho}\, \rho^{\alpha+\sigma}\, e^{-\rho}\, L_m^{(\alpha)}(\rho) = \frac{(-)^m}{m!}\, \MeijerG{y}{0,-\sigma,-\alpha-\sigma}{}{0}{m-\sigma}
\end{equation}
\begin{proof}
Substituting the Mellin-Barnes representation for $(1+y\rho)^{-1}$ of equation \eqref{eq:mbratio} in the
expression above, interchanging the integrals over $\rho$ and $t$, and using the well-known
integral, e.g. see 7.14-11 in \cite{Gra65},
\begin{equation}
\int_0^\infty \rho^{\alpha+\sigma+1+t}\, e^{-\rho}\, L_m^{(\alpha)}(\rho) d\rho = \frac{1}{m!}\Gamma(\alpha+\sigma+2+t) \frac{\Gamma(m-1-\sigma-t)}{\Gamma(-1-\sigma-t)}
\label{eq:intlag}
\end{equation}
we find that
\begin{eqnarray}
\int_0^\infty \frac{d\rho}{1+y\rho}\, \rho^{\alpha+\sigma}\, e^{-\rho}\, L_m^{(\alpha)}(\rho) &=&
  -\frac{1}{m!}\frac{1}{2\pi i}\int_{\gamma-i\infty}^{\gamma+i\infty}dt\;y^{1+t}\Gamma(1+t)\Gamma(-t) \Gamma(\alpha+\sigma+2+t) \frac{\Gamma(m-1-\sigma-t)}{\Gamma(-1-\sigma-t)} \nonumber\\
&=& \frac{1}{m!}\frac{1}{2\pi i}\int_{\gamma-i\infty}^{\gamma+i\infty}dt\;y^{1+t}\Gamma(2+t)\Gamma(-1-t) \Gamma(\alpha+\sigma+2+t) \frac{\Gamma(m-1-\sigma-t)}{\Gamma(-1-\sigma-t)} \nonumber\\
&=& \frac{(-)^m}{m!}\, \MeijerG{y}{0,-\sigma,-\alpha-\sigma}{}{0}{m-\sigma}
\label{eq:intcomplex}
\end{eqnarray}
where the last equality follows directly from the definition of the Meijer G-function,
and the step before follows from the identity $\Gamma(1+t)\Gamma(-t) = -\Gamma(2+t)\Gamma(-1-t)$.
\end{proof}
\end{lemma}
We also need the following sum over the Laguerre polynomial, combined with factorials
and rationals:
\begin{lemma}
\label{lem:sum}
For complex $\alpha,\beta,\lambda$, with $\Re\beta<\Re\alpha$ and valid for all $\rho$ we have
\begin{equation}
\sum_{n=0}^\infty \frac{(\beta+1)_n}{(\alpha+1)_n}\, \frac{L_n^{(\alpha)}(\rho)}{n+\lambda} =
\frac{\Gamma(\alpha+1)}{\Gamma(\beta+1)\Gamma(\lambda-\beta)}\, \MeijerG{\rho}{1-\lambda,-\beta}{}{-\beta,0}{-\alpha}
\label{eq:complexsum}
\end{equation}
\begin{proof}
Expressing the Laguerre polynomial in Kummer's function ${}_1F_1(\dots)$ and writing the sum
over $n$ as a contour integral we get for the sum \eqref{eq:complexsum}
\begin{equation}
\frac{1}{2\pi i}\int_\Gamma du\, (-1)^u\, \frac{\Gamma(-u)\Gamma(\beta+1+u)}{\Gamma(\beta+1) (u+\lambda)}\, {}_1F_1(-u;\alpha+1;\rho)
\end{equation}
Substituting for the Kummer function its Mellin-Barnes contour integral representation,
performing the contour integration over the variable $u$ the result follows immediately
from the definition of the Meijer G-function, e.g. see 9.301 in \cite{Gra65}.
\end{proof}
\end{lemma}
This lemma lets us prove the following significant result involving the operator $Q_\rho$:
\begin{lemma}
When the parameter $\lambda = \tau-z-s$ we have:
\begin{eqnarray}
\int_0^\infty d\rho'\, G(\rho,\rho')\, Q_{\rho'}\, \rho'^\sigma &=& -\left( \frac{1}{2} + \left(\sigma+\tau-1-s+\ldots\right)y \right) \cdot\nonumber \\*
&&\cdot \int_{\cal L} \frac{du}{2\pi i}\, \rho^u\, \frac{\Gamma(-u)\Gamma(\lambda+u)}{\Gamma(\alpha+1+u)}\, \MeijerG{y}{0,-\alpha-\sigma,u-\sigma}{-\sigma}{0,u-\sigma}{-\lambda-\sigma} \nonumber \\
\label{eq:opqpower}
\end{eqnarray}
where the contour $\cal L$ is such that the poles of $\Gamma(-u)$, $\Gamma(1+\sigma-u)$
are to the right and the poles of $\Gamma(\lambda+u)$, $\Gamma(u-\sigma)$ are to the left
of the contour.
\begin{proof}
From the expression for $Q_\rho$ in \eqref{eq:opQ} we find that the action on a power of $\rho$ is
\begin{equation}
Q_\rho \rho^\sigma = \frac{1}{1+y\rho} \rho^\sigma \left( \frac{1}{2} + y \left(\sigma+\tau-1-s+\ldots\right) \right)
\end{equation}
and applying \eqref{eq:mbratio}, \eqref{eq:intlag}, \eqref{eq:intcomplex}, \lemref{lem:sum}, and
reordering the contour integral stemming from the Meijer G-functions, the result \eqref{eq:opqpower} follows.
\end{proof}
\end{lemma}
From the second lemma, it follows that for $\rho\sim 0$:
\begin{corollary}
When the parameter $\lambda = \tau-z-s$ and $\sigma$ is equal to a positive integer $n\geq0$:
\begin{equation}
\int_0^\infty d\rho'\, G(\rho,\rho')\, Q_{\rho'}\, \rho'^n = \sum_{m=0}^\infty a_{n,m}(y)\, \rho^m
\label{eq:opQrhozero}
\end{equation}
and
\[ a_{n,0} = -\left( \frac{1}{2} + \left(n+\tau-1-s+\ldots\right)y \right)\, \frac{\Gamma(\lambda)}{\Gamma(\alpha+1)}\, \MeijerG{y}{0,-n,-\alpha-n}{}{0}{-\lambda-n} \]
Proof. The result follows from a straightforward application of the residue theorem
to \eqref{eq:opqpower}.
\end{corollary}
Likewise for $\rho>>1$ we have the result:
\begin{corollary}
When the parameter $\lambda = \tau-z-s$ and $\sigma = -\lambda - n$ for a positive integer
$n\geq0$ we have:
\begin{equation}
\int_0^\infty d\rho'\, G(\rho,\rho')\, Q_{\rho'}\, \rho'^{-\lambda-n} = \rho^{z-\tau+s}\, \sum_{m=0}^\infty b_{n,m}(y)\, \rho^{-m}
\label{eq:opQrhoinfty}
\end{equation}
and
\[ b_{n,0} = -\left( \frac{1}{2} + \left(z-n-1+\ldots\right)y \right)\, \frac{\Gamma(\lambda)}{\Gamma(\alpha+1)}\, \MeijerG{y}{0,\lambda-\alpha+n,\lambda+n}{}{0}{n} \]
Proof. The result follows from a straightforward application of the residue theorem
to \eqref{eq:opqpower}.
\end{corollary}
Finally, we are in a position to prove the following theorem and main result of this section:
\begin{theorem}
The integral equation
\begin{equation}
k(\rho) = \int_0^\infty d\rho' \, G(\rho,\rho') Q_{\rho'} k(\rho')
\label{eq:integraleqn2}
\end{equation}
is a solution to \eqref{eq:basicform} satisfying the boundary conditions
\eqref{eq:boundarycond1}-\eqref{eq:boundarycond3}.
\begin{proof}
We start from equation \eqref{eq:homplusint} and examine it near
$\rho=0$ (a) and $\rho=\infty$ (b). Notice that the parameter $\lambda$ is arbitrary
and we make the choice $\lambda=\tau-z-s$. For this choice the operator $Q_\rho$ \eqref{eq:opQ}
reduces to
\[ Q_\rho = \frac{1}{1+y\rho}\, \left( \frac{1}{2} + y(\tau-1-s+\ldots) + y\rho\,\frac{d}{d\rho} \right) \]
and
\[ Q_\rho\, \rho^\sigma = \frac{\rho^\sigma}{1+y\rho}\, \left( \frac{1}{2} + y(\sigma+\tau-1-s+\ldots) \right) \]
\paragraph{Boundary condition at $\rho=0$~\eqref{eq:boundarycond1}.} We know that the Frobenius
solution near $\rho=0$ is:
\[ k(\rho) = \sum_{n=0}^\infty k_{0,n}\, \rho^n \]
Substituting into \eqref{eq:homplusint} and using \eqref{eq:opQrhozero} gives:
\begin{eqnarray}
k(\rho) &=& C_1\, M(\lambda;\alpha+1;\rho) + C_2\, U(\lambda;\alpha+1;\rho) + \sum_{n=0}^\infty k_{0,n}\, \int_0^\infty d\rho' \, G(\rho,\rho') Q_{\rho'} \rho'^n \nonumber\\
&=& C_1 {}M(\lambda;\alpha+1;\rho) + C_2\, U(\lambda;\alpha+1;\rho) + \sum_{m=0}^\infty \rho^m \left[ \sum_{n=0}^\infty k_{0,n}\, a_{n,m}(y) \right]
\label{eq:rhozero}
\end{eqnarray}
Given that $U(\lambda;\alpha+1;\rho)\sim\rho^{-\alpha}$ for small $\rho$ and considering the lowest
power of $\rho$ in \eqref{eq:rhozero} it follows that the constant $C_2=0$.\\
\paragraph{Boundary condition at $\rho=\infty$~\eqref{eq:boundarycond3}.} Since we know that a
normal solution at $\rho=\infty$ exists and is of the form:
\[ k(\rho) = \rho^{z-\tau+s}\, \sum_{n=0}^\infty k_{\infty,n}\, \rho^{-n} \]
Substituting into \eqref{eq:homplusint} and using \eqref{eq:opQrhoinfty} together with (13.1.4)
in \cite{AS70} gives:
\begin{eqnarray}
k(\rho) &=& C_1\, M(\lambda;\alpha+1;\rho) + C_2\, U(\lambda;\alpha+1;\rho) + \sum_{n=0}^\infty k_{\infty,n}\, \int_0^\infty d\rho' \, G(\rho,\rho') Q_{\rho'} \rho'^{z-\tau+s-n} \nonumber\\
&=& C_1\, \frac{\Gamma(\alpha+1)}{\Gamma(\lambda)}\, e^\rho\, \rho^{-z-s-\alpha} (1+{\cal O}(1/\rho)) + C_2\, \rho^{z-1+s}\, (1+{\cal O}(1/\rho)) \nonumber\\
&& + \rho^{z-\tau+s}\, \sum_{m=0}^\infty \rho^{-m} \left[ \sum_{n=0}^\infty k_{\infty,n}\, b_{n,m}(y) \right]
\label{eq:rhoinfinity}
\end{eqnarray}
Again, given the exponential behaviour, we see that $C_1=0$.
This completes the proof.
\end{proof}
\end{theorem}

%
\section{\label{sec:shatproperties}The special functions $\bm{S^m_n}$ and $\hat{\bm S}^{\bm m}_{\bm n}$}
This appendix introduces the family of $S$-functions that occur in the determinant \eqref{eq:detprime}.
\subsection{Definition}
\begin{definition}
The functions $\overline{\bm S}^m_n$ and $\hat{\bm S}^m_n$ are defined as
\begin{eqnarray}
\overline{S}^m_n &=& \frac{m!}{\Gamma ( \alpha +m+1)} \sum _{k=0}^{n} I_{m,k} = \overline{I}_{m,n}^{(\alpha,\alpha+1)} \label{eq:defsmn}\\
\shat^m_n &=& \overline{S}_{n}^{m} - \omega y\; \overline{I}_{m,n} \label{eq:defshat}
\end{eqnarray}
where we have used the property of Laguerre polynomials $\sum_{k=0}^n L_k^{(\alpha)}(z) = L_n^{(\alpha+1)}(z)$
and $\omega$ is defined as
\[ \omega = \frac{1}{2y} - \tau + s + 1 - \ldots \]
\end{definition}
%
\subsection{Representation}
An explicit representation of $\shat^m_n$ is:
\begin{equation}
\begin{cases}
\displaystyle 1-\shat_{n}^{m}(y) = \frac{m!}{( \alpha+1)_m} \ L_{m}^{( \alpha )}( -1/y) \ \left( y(\alpha+1)\ \overline{I}_{0,n}^{( \alpha +1 )}( y) + \omega y\; \overline{I}_{0,n}^{( \alpha )}( y)\right) & m\leqslant n\\
 & \\
\displaystyle \shat_{n}^{m}(y) = \overline{I}_{m,0}^{( \alpha )}( y) \ \left( L_{n}^{( \alpha +1)}( -1/y) - \omega y\; L_{n}^{( \alpha )}( -1/y)\right) & m >n
\end{cases}
\label{eq:shatrepr}
\end{equation}
As a special case, we have
\[ \shat_0^m(y) = (1 - \omega\ y)\; \overline{I}_{m,0}^{( \alpha )}( y) \]
The importance of the representation \eqref{eq:shatrepr} is that it shows that, depending on whether
$m>n$ or $n\leq n$ the $\shat$ factorizes into two components, each containing the $m$ or $n$ dependence.
This fact is exploited in its symmetry properties.
%
\subsection{Symmetry properties}
From the representation \eqref{eq:shatrepr} it directly follows that:
\begin{eqnarray}
&&\shat_{n}^{m} \ \shat_{\ell }^{k} - \shat_{n}^{k} \ \shat_{\ell }^{m} = 0, \text{when $m,k\geq n,\ell$} \label{eq:shatsymmetry} \\
&&\left( 1-\shat_{n}^{m}\right) \ \left( 1-\shat_{\ell }^{k}\right) - \left( 1-\shat_{n}^{k}\right) \ \left( 1-\shat_{\ell }^{m}\right) = 0, \text{when $m,k\leq n+1, \ell+1$} \label{eq:shatasymmetry}
\end{eqnarray}
%
\subsection{Asymptotic behavior}
The asymptotic behavior of the function $\sbar^m_n$ defined in \eqref{eq:defsmn} is given by the
next lemma.
\begin{lemma}
The function $\sbar^m_n$ has the asymptotic behavior
\begin{equation}
\begin{cases}
\displaystyle 1-\sbar^m_n \sim \sqrt{\pi}\, (\alpha+1)\; \frac{m!}{(\alpha+1)_m}\; L_m^{(\alpha)}(-1/y)\; y^{-\frac{1}{2}\alpha-\frac{1}{4}}\, e^{\frac{1}{2y}}\, n^{\frac{1}{2}\alpha+\frac{1}{4}}\, e^{-2\sqrt{n/y}}, & \mbox{with $m$ fixed and $n\to\infty$} \\[1em]
\displaystyle \sbar^m_n \sim \sqrt{\pi}\, y^{-\frac{1}{2}\alpha-\frac{3}{4}}\, e^{\frac{1}{2y}}\, L_n^{(\alpha+1)}(-1/y)\, m^{-\frac{1}{2}\alpha-\frac{1}{4}}\, e^{-2\sqrt{m/y}},& \mbox{with $n$ fixed and $m\to\infty$} \\[1em]
\displaystyle \sbar^m_m \sim \frac{1}{2},& \mbox{when $m=n$ and $m,n\to\infty$},
\end{cases}
\end{equation}
\begin{proof}
This follows from its definition \eqref{eq:defsmn}, \eqref{eq:iaa1}, and \thmref{lem:asymptotic}, and
\end{proof}
\end{lemma}
\begin{corollary}
The asymptotic behavior for $\shat^m_n$ defined in \eqref{eq:defshat} is:
\begin{equation}
\begin{cases}
\displaystyle 1-\shat^m_n \sim \sqrt{\pi}\, (\alpha+1)\; \frac{m!}{(\alpha+1)_m}\; L_m^{(\alpha)}(-1/y)\; y^{-\frac{1}{2}\alpha-\frac{1}{4}}\, e^{\frac{1}{2y}}\, n^{\frac{1}{2}\alpha+\frac{1}{4}}\, e^{-2\sqrt{n/y}}, &\mbox{with $m$ fixed and $n\to\infty$} \\[1em]
\displaystyle \shat^m_n \sim \left( L_n^{(\alpha+1)}(-1/y) - \omega\ y\, L_n^{(\alpha)}(-1/y)\right)\sqrt{\pi}\, y^{-\frac{1}{2}\alpha-\frac{3}{4}}e^{\frac{1}{2y}}\, m^{-\frac{1}{2}\alpha-\frac{1}{4}}\, e^{-2\sqrt{m/y}}, &\mbox{with $n$ fixed and $m\to\infty$} \\[1em]
\displaystyle \shat^m_n \sim \frac{1}{2},\,\, &\mbox{when $m=n$ and $m,n\to\infty$}
\end{cases}
\label{eq:stildeasymp}
\end{equation}
\end{corollary}
\begin{corollary}
From the definition of $\shat$ and \eqref{eq:imnasymp} it is easily derived that for large $y$ and fixed $m,n$:
\begin{eqnarray}
\label{eq:shatlargey}
\shat_{n}^{m} &\sim& \frac{m!}{( \alpha +1)_{m}}\ \frac{( \alpha +1)_{n}}{n!} \ \frac{1}{\alpha\ y} \ \left(\frac{n}{\alpha +1} + 1 - y\;\omega\right) ,\ m\geqslant n \\
\shat_{n}^{m} &\sim& \left( n-m+1 + \frac{m}{\alpha+1} - y\;\omega \right)\; \frac{1}{\alpha y} ,\ m\leqslant n
\end{eqnarray}
\end{corollary}
The behaviour of the $\shat$ for large indices around the diagonal of the determinant \eqref{eq:detexplicit}
is described in the following theorem. The results follow immediately from the definition of the $\shat$
\eqref{eq:defshat}, expansion from the ratio of gamma-functions \cite[\href{https://dlmf.nist.gov/5.11.iii}{(5.11(iii))}]{DLMF},
expansion of Kummer's $U$-function \cite{Tem13}, and Laguerre polynomials \cite{Bor07}. For the evaluation and
manipulation of the algebraic expressions, we have used the computer algebra program FORM \cite{Ver00}.
\begin{theorem}
For $\tau=1$, $\ldots=0$ the $\shat$ and $\shat^{(\alpha-1)}$ have asymptotic expansion in $y/n$ for the diagonal and
off-diagonal elements where $n>>y$:
\begin{subequations}
\begin{align}
\shat_n^n &= \frac{1}{2} - \frac{5}{8}\ \sqrt{\frac{y}{n}}
{}+ \left(-\frac{7}{256} + \frac{7}{64}\alpha^2 + \frac{\alpha+1}{64y} \left( 9 + 4\alpha^2 \right) + \frac{3}{64}\frac{1}{y^2} \right) \left(\frac{y}{n}\right)^{\frac{3}{2}}
{}+ \cdots \\[1em]
\shat_{n-1}^{n-1} &= \frac{1}{2} - \frac{5}{8}\ \sqrt{\frac{y}{n}}
{}+ \left(-\frac{7}{256} + \frac{7}{64}\alpha^2 + \frac{1}{64y} \left( 9\alpha - 11 + 4\alpha^2(\alpha+1) \right) + \frac{3}{64}\frac{1}{y^2} \right) \left(\frac{y}{n}\right)^{\frac{3}{2}}
{}+ \cdots \\[1em]
\shat_n^{n-1} &= \frac{1}{2} - \left(\frac{5}{8} - \frac{1}{2y}\right) \sqrt{\frac{y}{n}} + \frac{1}{4y}\ \left(2-\alpha - \frac{1}{y}\right)\frac{y}{n}
{}+\left( -\frac{7}{256} + \frac{7}{64}\alpha^2 + \frac{1}{64y} (-2 - 11\alpha + 8\alpha^2 + 4\alpha^3)\right. \nonumber\\[1em]
&\left.+ \frac{1}{64y^2} (-9 + 8\alpha) + \frac{1}{16y^3} \right) \left(\frac{y}{n}\right)^{\frac{3}{2}}
{}+ \cdots \\[1em]
\shat_{n-1}^n &= \frac{1}{2} - \left(\frac{5}{8} + \frac{1}{2y}\right) \sqrt{\frac{y}{n}} + \frac{1}{4y}\ \left(2-\alpha + \frac{1}{y}\right) \frac{y}{n}
{}+\left(-\frac{7}{256} + \frac{7}{64}\alpha^2 + \frac{1}{64y}\left( -5 + 11\alpha - 20\alpha^2 - 4\alpha^3 \right) \right.\nonumber\\[1em]
&\left.+ \frac{3}{64y^2} (- 3 + 8\alpha ) - \frac{1}{16}\frac{1}{y^3} \right) \left(\frac{y}{n}\right)^{\frac{3}{2}}
{}+ \cdots \\[1em]
\shat_n^{(\alpha-1),n} &= \frac{1}{2} - \frac{3}{8}\ \sqrt{\frac{y}{n}} + \left( \frac{15}{256} - \frac{5}{32} \alpha + \frac{5}{64} \alpha^{2}
{}+ \frac{\alpha}{64y} \left( 9 - 8\alpha + 4\alpha^{2} \right) + \frac{1}{64} \frac{1}{y^{2}} \right)\left(\frac{y}{n}\right)^{\frac{3}{2}} + \cdots \\[1em]
\shat_{n-1}^{(\alpha-1),n-1} &= \frac{1}{2} - \frac{3}{8}\ \sqrt{\frac{y}{n}} + \left(\frac{15}{256} - \frac{5}{32} \alpha + \frac{5}{64} \alpha^{2}
{}+ \frac{1}{64y} \left( -12 + 9\alpha- 8\alpha^{2} + 4\alpha^{3} \right) + \frac{1}{64} \frac{1}{y^{2}}\right) \left(\frac{y}{n}\right)^{\frac{3}{2}} + \cdots \\[1em]
\shat_n^{(\alpha-1),n-1} &= \frac{1}{2} - \left(\frac{3}{8} - \frac{1}{2y}\right) \sqrt{\frac{y}{n}}
{}- \frac{1}{4y}\left( 2-\alpha - \frac{1}{y} \right) \frac{y}{n}
{}+ \left( \frac{15}{256} - \frac{5}{32} \alpha + \frac{5}{64} \alpha^{2} + \frac{1}{64y}\left( 10 - 13 \alpha - 4 \alpha^{3}\right)\right. \\[1em]
&\left. + \frac{1}{64y^{2}}\left( -11 + 8 \alpha\right) + \frac{1}{16} \frac{1}{y^3}
\right)\left(\frac{y}{n}\right)^{\frac{3}{2}} + \cdots\\[1em]
\shat_{n-1}^{(\alpha-1),n} &= \frac{1}{2} - \left(\frac{3}{8} + \frac{1}{2y}\right) \sqrt{\frac{y}{n}}
{}+ \frac{1}{4y}\left( 2-\alpha + \frac{1}{y} \right) \frac{y}{n}
{}+ \left( \frac{15}{256} - \frac{5}{32} \alpha + \frac{5}{64} \alpha^{2} + \frac{1}{64y}\left( - 20 + 27 \alpha - 8 \alpha^{2} - 4 \alpha^{3}\right)\right. \\[1em]
&\left. + \frac{3}{64y^{2}}\left( -9 + 8 \alpha\right) - \frac{1}{16} \frac{1}{y^3}
\right)\left(\frac{y}{n}\right)^{\frac{3}{2}} + \cdots \\[1em]
1 - \shat^{0}_n &= \frac{\sqrt{\pi}}{\Gamma(\alpha+1)} \left(\frac{n}{y}\right)^{\frac{1}{2}\alpha+\frac{1}{4}}\ e^{-2\sqrt{\frac{n}{y}}} \left(1 + \frac{1}{2y} + \frac{1}{8y^2} + \frac{1}{48y^{3}}
{}+ \left( \frac{19}{16} + \frac{1}{4} \alpha^2
{}+ \frac{1}{32y}\left( 3 - 16 \alpha + 4 \alpha^2\right) \right.\right. \\[1em]
&\left.\left.
{}+ \frac{1}{64y^{2}}\left( -\frac{71}{6} - 16 \alpha + 2 \alpha^2\right)
{}+ \frac{1}{192y^{3}}\left( -\frac{61}{4} - 12 \alpha + \alpha^2\right)
{}+ \frac{1}{96y^{4}}\left( -2 - \alpha\right) - \frac{1}{576} \frac{1}{y^5}\right) \sqrt{\frac{y}{n}} + \cdots \right) \\[1em]
1 - \shat^{(\alpha-1),0}_n &= \frac{\sqrt{\pi}}{\Gamma(\alpha)} \left(\frac{n}{y}\right)^{\frac{1}{2}\alpha-\frac{1}{4}}\ e^{-2\sqrt{\frac{n}{y}}} \left(1 + \frac{1}{2y} + \frac{1}{8y^2} + \frac{1}{48y^{3}}
{}+ \left( \frac{15}{16} - \frac{1}{2}\alpha + \frac{1}{4} \alpha^2
{}+ \frac{1}{32y}\left( 15 - 24 \alpha + 4 \alpha^2\right) \right.\right. \\[1em]
&\left.\left.
{}+ \frac{1}{64y^{2}}\left( \frac{13}{6} - 20 \alpha + 2 \alpha^2\right)
{}+ \frac{1}{192y^{3}}\left( -\frac{17}{4} - 14 \alpha + \alpha^2\right)
{}+ \frac{1}{96y^{4}}\left( -1 - \alpha\right) - \frac{1}{576} \frac{1}{y^5}\right) \sqrt{\frac{y}{n}} + \cdots \right)
\end{align}
\end{subequations}
\end{theorem}
\begin{figure}[htb]
\begin{center}
\begin{subfigure}{0.45\textwidth}
\includegraphics[width=\textwidth]{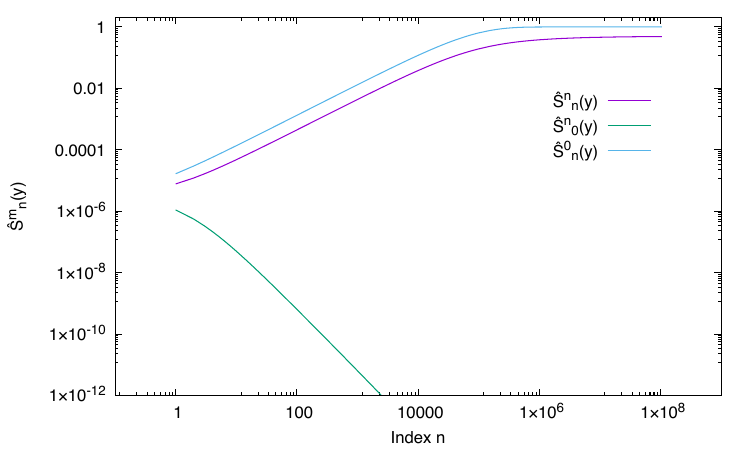}
\caption{The $\shat^m_n$-function (see \eqref{eq:shatlargey} and \eqref{eq:stildeasymp}).\label{fig:shatbasic}}
\end{subfigure}
\begin{subfigure}{0.45\textwidth}
\includegraphics[width=\textwidth]{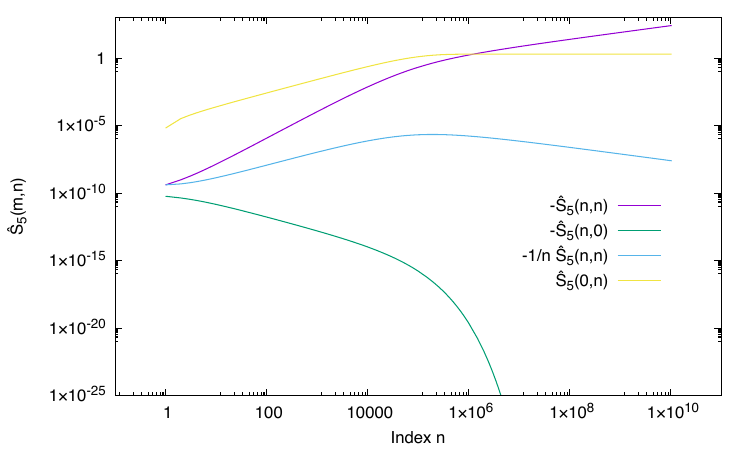}
\caption{The $\shat_{5}(m,n)$-function (see \eqref{eq:defshat5}).\label{fig:shat5basic}}
\end{subfigure}
\end{center}
\caption{The $\shat$- and $\shat_{5}$-functions with varying index $n$ illustrating the
"small" region ($n<<y$) as well as the asymptotic region $y>>n$.\label{fig:shatbasics}}
\end{figure}
\subsection{\label{sec:combinations}Combinations that are of order two or higher}
In performing the calculations for the second-order contribution to the determinant, we have found it useful
to have built-in checks. One such verification is to have each term in the expansion of the determinant
to be explicitly of second order. The $\shat$ are of first order and certain combinations exist in which
the leading asymptotic terms, for large $y$, cancel, giving second order terms. A list of combinations
that exhibit this asymptotic property is summarized in the next theorem.
\begin{theorem}
Define the combinations of $\shat^m_n$:
\begin{subequations}
\label{eq:combinations}
\begin{eqnarray}
\shat_{0} &=& \shat^{0}_{0} - \frac{1}{4} \frac{1}{\alpha y} \\[1em]
\shat_{1}(n) &=& \shat^{n}_{n} - \shat^{n-1}_{n-1} \label{eq:defshat1}\\[1em]
\shat_{2}(n) &=& \alpha \left( \shat^{n}_{n} - \shat^{n-1}_{n-1} \right) + \left( \shat^{n-1}_{n-1} - \shat^{n-2}_{n-1} \right) \\[1em]
\shat_{3}(n) &=& \shat^{n}_{n} - \shat^{n-1}_{n-1} - \shat^{n-1}_{n} + \shat^{n-2}_{n-1} \\[1em]
\shat_{4}(n) &=& \shat^{0}_{n} - \shat^{0}_{n-1} - \frac{1}{\alpha y} \label{eq:defshat4}\\[1em]
\shat_{5}(n,m) &=& (n+\alpha+1) \ \shat^{n+1}_{m} - \left(n+1\right) \ \shat^{n}_{m} \label{eq:defshat5}\\[1em]
\shat_{6}(n) &=& \alpha \left( \shat^{n}_{n} - \shat^{n-1}_{n-1} \right) + \left( \shat^{n}_{n} - \shat^{n-1}_{n} \right) = \shat_2(n) + \shat_3(n) \\[1em]
\shat_{7}(n,m) &=& (n+\alpha) \ \shat_{5}(n,m) - (n+1) \ \shat_{5}(n-1,m)
\end{eqnarray}
\end{subequations}
then each combination is of order ${\cal O}(1/y^2)$, except for $\shat_1$ which is of order ${\cal O}(1/y)$ and $\shat_{7}$ which is of order ${\cal O}(1/y^3)$.
\begin{proof}
The result follows from the definition \eqref{eq:defshat} and the asymptotic expansion of the $I_{0,n}$ \eqref{eq:imnasymp}.
\end{proof}
\end{theorem}
%
\subsection{Recurrence relations}
Relations between consecutive $\shat$-functions are straightforward to derive from the
definition \eqref{eq:defsmn} and the recurrence relations between the $I_{m,n}$-functions
\eqref{eq:ttrr}.
\begin{eqnarray*}
m\ \Big( \shat^{m+1}_{m} - 2\shat^{m}_{ m} + \shat^{m-1}_{m} \Big) &=& (\alpha+1) \left(\shat^{m}_{ m} - \shat^{m+1}_{ m} \right) + \frac{1}{y} \ \shat^{m}_{ m} - (1-\omega\ y) \frac{1}{y}, \mbox{ for $m=n$}\\
m\ \Big(\shat^{m+1}_{n} - 2\shat^{m}_{n} + \shat^{m-1}_{n}\Big) &=& (\alpha+1)\ (\shat^{m}_{n} - \shat^{m+1}_{n}) + \frac{1}{y} \shat^{m}_{n}, \mbox{ for $m>n$} \\
m\ \Big(\shat^{m+1}_{n} - 2\shat^{m}_{n} + \shat^{m-1}_{n}\Big) &=& (\alpha+1)\ (\shat^{m}_{n} - \shat^{m+1}_{n}) + \frac{1}{y} \ \shat^{m}_{n} - \frac{1}{y}, \mbox{ for $m<n$}
\end{eqnarray*}
The special case for $m=0$ is addressed in the following relations:
\begin{eqnarray*}
(\alpha+1)\ \shat^{1}_{n} &=& (\alpha + 1 + 1/y)\ \shat^{0}_{n} -\frac{1}{y}, \mbox{ for $n>0$}\\
(\alpha+1)\, \shat^{1}_{0} &=& (\alpha + 1 + 1/y)\, \shat^{0}_{0} - \frac{1}{y} (1 - \omega\ y), \mbox{ for $n=0$}
\end{eqnarray*}
Similar relations for the combinations of $\shat$-functions are listed below:
\begin{eqnarray*}
\shat_{5}(n+1,n+1)-\shat_{5}(n,n) &=& \frac{1}{y} \shat_{1}(n+1) + \shat_{2}(n+1) + (n+1)\ \shat_{3}(n+1) \\
\shat_{5}(n,0) - \shat_{5}(n-1,0) &=& \frac{1}{y} \ \shat^{n}_0 \\
\shat_{5}(n,m) - \shat_{5}(n-1,m) &=& \frac{1}{y} \ \shat^{n}_{m}, \text{when $n>m$} \\
\shat_{5}(n,m) - \shat_{5}(n-1,m) &=& \frac{1}{y} \ \shat^{n}_{m} - \frac{1}{y}\, (1-\omega\ y)\, \delta_{m,n} \\
\shat_{7}(m+1,m) - \shat_{7}(m,m) &=& \frac{1}{y}\, \shat_{5}(m,m) + \frac{1}{y}\, (1-\omega\ y) (m+1)
\end{eqnarray*}
Besides the recurrence relations in the upper index, there exist a few relations involving the lower indices.
This can be used to convert a recurrence in the lower indices to one that is expressed in the upper index.
The side effect is that it also lowers the parameter $\alpha$. The relations are:
\begin{align}
\shat^{\alpha,n}_{m-1} - \shat^{\alpha,n}_{m-2} &= \shat^{\alpha-1,n}_{m-1} - \shat^{\alpha-1,n+1}_{m-1}, \text{for $n\geq m$} \\[1em]
\shat^{\alpha,m}_{m+1} - \shat^{\alpha,m}_{m} &= \shat^{(\alpha-1),m}_{m+1} - \shat^{(\alpha-1),m+1}_{m+1} \\[1em]
\shat_{5}(n-1,m-1) - \shat_{5}(n-1,m-2) &= -\frac{1}{y}\ \shat^{(\alpha-1),n}_{m-1}, \text{for $n\geq m$} \\[1em]
\shat_5(m-2,m-1) - \shat_5(m-2,m-2) &= -\frac{1}{y}\ \shat^{(\alpha-1),m-1}_{m-1} + \left(\frac{1}{y}-\omega\right) \\[1em]
\shat_5(n,m-1) - \shat_5(n,m-2) &= -\frac{1}{y}\ \shat^{(\alpha-1),n+1}_{m-1} + \frac{1}{y}, \text{for $n<m-2$}
\end{align}
\subsection{Other relations}
This section lists miscellaneous relations between products of $\shat$-functions.
\begin{eqnarray*}
\shat^{m-1}_0\ \shat^{m}_{ m} - \shat^{m-1}_{m}\ \shat^{m}_0 &=& (1-\omega\ y) \frac{1}{y}\ \frac{1}{m}\shat^{m}_0 \\[1em]
\shat^{m-2}_{m}\shat^{m}_0 - \shat^{m-2}_0\shat^{m}_{m} &=& - \frac{\alpha+2m-1+1/y}{m(m-1)} \left(1-\omega\ y\right) \frac{1}{y}\ \shat^{m}_0 - \frac{1}{y} \frac{1}{m-1} \shat^{m}_0 \\[1em]
\shat^{n}_{0} \left( \shat^{n+1}_{n} - \shat^{n}_{n} \right) &=& \frac{1}{n}\shat^{n}_{n} \left( \shat_{5}(n,0) - (\alpha+1)\shat^{n+1}_{0} + \shat^{n}_{0} \right) \\[1em]
\shat^{n}_{0}\, \shat_{5}(n,n) &=& \shat_{5}(n,0)\, \shat^{n}_{n} \\[1em]
\shat^{n+1}_{0}\, \shat_{5}(n,n) &=& \shat_{5}(n,0)\, \shat^{n+1}_{n} \\[1em]
\left(\shat_{n}^{m} -1\right) \ \shat_{n+1}^{n+1} -\shat_{n}^{n+1} \ \left(\shat_{n+1}^{m} -1\right) &=& - \frac{n+2-s-z_{y}}{\alpha +n+1}\, \bar{I}_{m,n+1},\ m\leqslant n, \text{and $z_y = (2-s)sy - \frac{1}{4y}$}
\end{eqnarray*}

%

\end{document}